\documentclass[aps,prb,pdflatex,twocolumn,superscriptaddress]{revtex4-2}
\usepackage{amsmath}
\usepackage{amssymb}
\usepackage{amsfonts}
\usepackage{bbm}
\usepackage{amsthm}
\usepackage{xcolor}
\usepackage{hyperref}
\usepackage{braket}

\theoremstyle{remark}

\newcommand{\be}{\begin{equation}}
\newcommand{\ee}{\end{equation}}
\newcommand{\bea}{\begin{equation}\begin{aligned}}
\newcommand{\eea}{\end{aligned}\end{equation}}
\newcommand{\mbf}[1]{\mathbf{#1}}
\newcommand{\Nf}{N_f}
\newcommand{\Vol}{N_{\text{cell}}}

\newcommand{\Tr}{{\rm Tr}}
\newcommand{\del}{\partial}
\newcommand{\al}{\alpha}
\DeclareRobustCommand{\App}[1]{App.~\ref{#1}}

\newcommand{\Eq}[1]{Eq. \ref{#1}}

\begin{document}


\title{Exact Stiffness and Dynamical Responses from Fock-Space Fragmentation}

\author{Jonah~Herzog-Arbeitman}
\affiliation{Department of Physics, Princeton University, Princeton, New Jersey 08544, USA}
\affiliation{Department of Physics, Massachusetts Institute of Technology, Cambridge MA 02139, USA}

\author{Eslam~Khalaf}
\affiliation{Department of Physics, Harvard University, Cambridge, Massachusetts 02138, USA}

\author{Zhaoyu~Han}
\email{zhan@fas.harvard.edu}
\affiliation{Department of Physics, Harvard University, Cambridge, Massachusetts 02138, USA}

\begin{abstract}
Exactly solvable quantum many-body models are rare, and even when their spectra are algebraically organized, dynamical responses generally remain difficult to obtain because they probe an extensive number of excited states. Here we show that quantum geometric nesting (QGN) models admit an unusually strong form of solvability rooted in \emph{Fock-space fragmentation}: excitations on top of the exact frustration-free ground states decouple into Krylov subspaces with a fixed number of particle and hole operators, and hence remain dynamically invariant. Exploiting this structure, we prove that the stiffness of the spontaneously broken continuous symmetry in QGN models is exactly equal to its variational value in the Gaussian manifold, confirming a conjecture from quantum many-body bootstrap~\cite{GaoHanKhalaf2026}. The proof shows that an infinitesimal phase twist couples the ground state only to the one-particle, one-hole fragment, which coincides with the tangent space of the ground state within the variational manifold, thereby making the variational curvature exact. More generally, perturbations whose action remains within a fixed Fock-space fragment have response functions determined exactly by the corresponding few-body sector, enabling exact access to quantities including static susceptibility, optical conductivity, dynamical structure factors, and single-particle Green's functions. 
\end{abstract}
\maketitle

\section{Introduction}
\label{sec:intro}

Exactly solvable models, though rare in strongly coupled quantum many-body physics, have long played pivotal roles in elucidating complex phenomena and providing controlled expansions around strongly interacting points of the phase diagram. Solvability itself comes in qualitatively different levels. At one extreme, commuting-projector Hamiltonians and Kitaev-type models admit explicit decompositions into sectors labeled by local integrals of motion, thereby providing access to the full many-body spectrum~\cite{Kitaev2006,LevinWen2005}. A second broad class, encompassing models solvable by the Bethe ansatz as well as the Haldane--Shastry, Calogero--Sutherland, and Richardson--Gaudin models, possesses closed algebraic structures that organize and substantially reduce the many-body spectral problem~\cite{Bethe1931,Haldane1988,Shastry1988,Calogero1971,Sutherland1971,Dukelsky2004}. A weaker but widely encountered notion of solvability arises in frustration-free parent Hamiltonians, including fractional quantum Hall pseudopotential Hamiltonians and Rokhsar--Kivelson models, where exact ground states can be constructed as common zero modes of local positive-semi-definite terms without a corresponding generic solution of the excitation spectrum ~\cite{Haldane1983,TrugmanKivelson1985,RokhsarKivelson1988,AKLT1987}. For frustration free models without additional spectral structure, exact statements about excitations are typically limited to bounds: variational constructions such as the single-mode approximation yield rigorous upper bounds on excitation energies~\cite{GirvinMacDonaldPlatzman1985}, while quantum many-body bootstrap methods can provide complementary rigorous lower bounds on gaps, susceptibilities, and response coefficients~\cite{NancarrowXin2023,Rai2026hierarchyofspectral,GaoHanKhalaf2026}. Importantly, even when the spectrum is algebraically organized, as in Bethe-integrable models, dynamical response functions generally remain difficult to obtain: a local probe couples to a macroscopic set of excited states, requiring nontrivial form factors and the resummation of large Lehmann expansions~\cite{Caux2009,MurgKorepinVerstraete2012}. 

Quantum geometric nesting (QGN) models~\cite{QGN}, which provide idealized strong-coupling descriptions of symmetry-breaking orders in fermionic flat bands, occupy an unusual position in this hierarchy. They are frustration-free, yet their solvability extends far beyond the ground state because of a special algebraic structure. A QGN model can be generically written as
\begin{align}
    H=\sum_\alpha S_\alpha^\dagger S_\alpha
    ,
    \qquad
    S_\alpha^\dagger\ket{\mathrm{GS}}=S_\alpha\ket{\mathrm{GS}}=0,
\end{align}
where each \(S_\alpha\) is a fermion bilinear. Consequently, for any excitation operator \(\mathcal O\),
\begin{align}
    H\mathcal O\ket{\mathrm{GS}}
    &=[H,\mathcal O]\ket{\mathrm{GS}}
    =\sum_\alpha
    [S^\dagger_\alpha,[S_\alpha,\mathcal O]]
    \ket{\mathrm{GS}}.
    \label{eq:double_comm_intro}
\end{align}
Since commutation with a fermion bilinear preserves the numbers of fermion creation and annihilation operators in an operator of definite particle--hole composition, Eq.~\eqref{eq:double_comm_intro} implies an exact decoupling of the excitation spectrum. In particular, defining
\begin{align}
    \mathcal K_{p,h}
    =
    \operatorname{span}
    \left\{
    c^\dagger_{a_1}\cdots c^\dagger_{a_p}
    c_{b_1}\cdots c_{b_h}
    \ket{\mathrm{GS}}
    \right\},
\end{align}
one finds
\begin{align}
    H\mathcal K_{p,h}\subseteq\mathcal K_{p,h}.
\end{align}
Thus the Krylov spaces generated from the exact ground state are graded by the Fock-space composition \((p,h)\) of the excitations and cannot hybridize with sectors of different composition. We refer to this structure as \emph{Fock-space fragmentation}. The double-commutator structure sharply distinguishes QGN models from generic frustration-free Hamiltonians. For example, a general parent Hamiltonian takes the form \(H=\sum_\alpha L_\alpha^\dagger L_\alpha\), where \(L_\alpha\ket{\mathrm{GS}}=0\) but \(L_\alpha^\dagger\ket{\mathrm{GS}}\neq0\) in general, as in fractional quantum Hall pseudopotential Hamiltonians\cite{Haldane1983,TrugmanKivelson1985}. In that case, $
    H\mathcal O\ket{\mathrm{GS}}$ can at most be simplified to $\sum_\alpha L_\alpha^\dagger[L_\alpha,\mathcal O]
    \ket{\mathrm{GS}}$,  which need not preserve the particle--hole composition of \(\mathcal O\).

This Fock-space closure was first encountered in studies of exact excitations in interacting twisted bilayer graphene, where fixed particle--hole sectors were used to construct exact charged and neutral excitations~\cite{TBGV}, and was subsequently exploited to study three-body trion and spin-polaron excitations~\cite{SchindlerTrion2022,KhalafVishwanath2022}. Later, in the context of flat band superconductivity \cite{PT15,TPB22}, this property was used to exactly solve the Cooper pair excitations \cite{HA22} in the frustration-free Hubbard models discovered in Ref. \cite{Tovmasyan2016}. Similar results have been derived for magnetism in repulsive Hubbard models \cite{2024arXiv240207171K,2026arXiv260301922K}.

QGN reveals Fock-space fragmentation as a general consequence of a broader class of frustration-free flat-band Hamiltonians. Unlike ``classical'' fragmentation, which can be resolved in a local product-state basis~\cite{DeTomasiHetterichSalaPollmann2019,KhemaniHermeleNandkishore2020,MoudgalyaMotrunich2022}, the QGN fragments are defined relative to a many-body ground state with intrinsic real-space entanglement and are themselves generically spanned by entangled states. In this sense, QGN realizes an intrinsically quantum form of fragmentation~\cite{PhysRevB.110.165109,HanHartKhudorozhkovNandkishore2026}. For any fixed \(p+h\), the dimension of \(\mathcal{K}_{p,h}\) grows only polynomially with system size, making all fixed-rank few-body excitation problems tractable with polynomial scaling. This unusual structure naturally raises a more ambitious question: Can dynamical response functions, which generically require information about an extensive set of excited states, also become exactly accessible?

In flat-band systems, a particularly natural testing ground for this question is the stiffness associated with a spontaneously broken continuous symmetry. For superconductors, the superfluid stiffness controls the rigidity of the condensate phase and is intimately tied to the quantum geometry of the underlying bands~\cite{PT15,Julku16,Tovmasyan2016,Liang2017,TPB22,Huhtinen22,HA22,2024PhRvL.132b6002C}. Closely related geometric effects arise in the spin stiffness of flat-band and quantum Hall ferromagnets~\cite{Sondhi1993,Moon1995,WuDasSarmaQAHF2020,WuDasSarma2020,RepellinDongZhangSenthil2020} and in the counterflow stiffness of exciton condensates~\cite{HuHyartPikulinRossi2022,VermaGuerciQueiroz2024}. In a strictly flat band, where the conventional kinetic contribution is quenched, the stiffness is generated entirely by interactions acting through the quantum geometry of the projected Hilbert space.

Determining the stiffness beyond mean field is nevertheless a genuine many-body response problem. From the Kohn formula, both diamagnetic and paramagnetic contributions appear -- the latter involving a spectral sum over excited states. Consequently, mean-field expressions generally provide only estimates, because the response can admix particle--hole excitations beyond the variational manifold of mean-field states, effectively dressing the collective carriers and changing their mass. Rigorous many-body results have therefore largely taken the form of upper bounds derived from the diamagnetic response, optical sum rules, or variational principles~\cite{PTR98,HVR19,VHR21,MaoChowdhury23,MaoChowdhury24}. The problem becomes particularly transparent -- but also particularly challenging -- after projection into an isolated perfectly flat band. Since the one-body kinetic current is quenched, the leading electromagnetic coupling is interaction-generated and the projected current operator is generically quartic in the fermions~\cite{MaoChowdhury23,MaoChowdhury24}. Numerically unbiased calculations of the stiffness are correspondingly scarce, relying mainly on quasi-one-dimensional DMRG~\cite{PhysRevB.105.024502,ChanGremeaudBatrouni2022Designer} or specially designed sign-problem-free two-dimensional quantum Monte Carlo models~\cite{PhysRevB.102.201112,PhysRevLett.130.226001,HAPeri22,ZhangSunLiPanMeng2022}.

QGN models offer a unique setting in which this many-body problem can be posed sharply. Recently, the superfluid stiffness was constrained from both sides~\cite{GaoHanKhalaf2026} in certain superconducting QGN models. A BCS variational calculation gives the rigorous upper bound
\begin{align}
    D_s
    \leq
    D_s^{\mathrm{BCS}}
    =
    \frac{N_f}{2}\nu(1-\nu)m_{\mathrm{pair}}^{-1},
    \label{eq:BCSbound_intro}
\end{align}
where \(\nu\) is the flat-band filling, \(N_f\) counts the flat bands including spin, and \(m_{\mathrm{pair}}\) is the exact mass of the two-particle bound state~\cite{TormaLiangPeotta2018,Huhtinen22,HA22}. Independently, reduced-density-matrix bootstrap yields rigorous \emph{lower} bounds on \(D_s\), which were found to saturate Eq.~\eqref{eq:BCSbound_intro} to numerical accuracy for a range of QGN interactions~\cite{GaoHanKhalaf2026}. On the basis of this striking saturation, Ref.~\cite{GaoHanKhalaf2026} conjectured that the true many-body stiffness is exactly
\begin{align}
    D_s = D_s^{\mathrm{BCS}},
    \label{eq:stiffness_conjecture_intro}
\end{align}
but a proof was left open.

In this paper, we prove this conjecture and show more generally that Fock-space fragmentation makes a broad class of dynamical response functions exactly solvable in QGN models. For the stiffness, the perturbation generated by an infinitesimal phase twist \(\mathbf A\) acts on the ground state entirely within the one-particle--one-hole fragment \(\mathcal{K}_{1,1}\). We show that the relevant subspace coincides precisely with the tangent space to the BCS variational manifold at the exact ground state. Consequently, the exact first-order deformation of the many-body ground state is completely contained within the BCS tangent space, and the variational curvature exhausts the full Kohn response. The BCS upper bound is therefore saturated, which proves Eq.~\eqref{eq:stiffness_conjecture_intro}. Physically, this means that a weak supercurrent in a QGN superconductor does not generate additional many-body dressing of the Cooper pairs beyond a deformation of their internal wavefunction: the two-particle pair mass directly controls the true many-body phase stiffness. 

More generally, any probe field with fixed Fock-space composition accesses \emph{only} the corresponding fragmented Krylov sector. Its Lehmann representation can therefore be reduced to the associated few-body spectral problem, allowing exact computation of frequency- and momentum-dependent quantities including the single-particle Green's function, dynamical structure factor, and optical conductivity as we discuss here.

We will focus our proof of exact stiffness for the superconducting case, which admits straightforward generalizations to ferromagnetism and exciton superfluids.

\section{Hamiltonian and Algebraic Structure}
\label{sec:setup}

To begin, we summarize the key results of the construction of frustration-free QGN Hamiltonians with exact ground states and solvable elementary excitations. 

\subsection{Frustration-Free QGN Superconductors}
\label{sec:models}

We consider multi-band tight-binding models with $\al = 1,\dots, N_{\rm orb}$ orbitals per unit cell (including internal flavors such as spin) on $\Vol$ unit cells, with $\Nf$ degenerate flat bands separated from all remote bands by a gap $\Delta$ much larger than the interaction scale. We work in a Hilbert space spanned by the flat band operators 
\bea
\gamma^\dagger_{\mbf k, n} &= \sum_\al c^\dag_{\mbf{k},\al}U_{\alpha n}(\mbf k)
\eea
and $n = 1,\dots, N_f < N_{orb}$. Here the vectors $U_n(\mbf{k})$ are the Bloch wavefunctions of the flat bands. They define the gauge-invariant flat band projector $P(\mbf{k}) = \sum_{n=1}^{\Nf} U_n(\mbf{k})U^\dag_n(\mbf{k})$ and $Q(\mbf{k}) = 1 - P(\mbf{k}) $ is its complement. 

Next we consider a projected interaction Hamiltonian in the form 
\be
 H = \sum_{\mbf q, IJ} V_{IJ}(\mbf q) S^{\dag}_{\mbf q, I} \,  S_{\mbf q,J}, \ V_{IJ}(\mbf q) \succeq 0
\ee
with $V_{IJ}(\mbf q) = V_{JI}(\mbf q)=  V_{IJ}(-\mbf q)$ (note we take $V \succeq 0$ as the strength of the attraction). $H$ is manifestly positive-semi definite, written in terms of the generalized ``spin" operators 
\bea
 S_{\mbf q, I} &= \frac{1}{\sqrt{\Vol}}\sum_{\mbf k mn}\,
\gamma^\dagger_{\mbf k + \mbf q, m}\, [S_I(\mbf k + \mbf q, \mbf k)]_{mn}\gamma_{\mbf k, n}- \bar{s}_{\mbf q,I} \ .
\label{eq:Sq}
\eea
where $\bar{s}_{\mbf q,I}$ is a constant offset. 
The simplest examples in this family include the attractive Hubbard model satisfying the uniform-pairing condition~\cite{HA22,Tovmasyan2016}, for which the \(S\) operators can be chosen as the local spin-\(S^z\) densities on each microscopic orbital. In the particle--hole channel, the simplest examples are flat-band ferromagnets, where the corresponding \(S\) operators can be taken to be projected density operators~\cite{TBGV,SchindlerTrion2022,KhalafVishwanath2022}.

The models $H$ are constructed to be frustration-free, meaning they have zero-energy ground states $\ket{0_N}$ which obey the local condition
\bea
S_{\mbf{q},I}\ket{0_N} &= 0, \qquad \forall \, \mbf{q},I \ . 
\eea
To construct the states $\ket{0_N}$, we will use a pairing operator (we assume $F(\mbf k) = -F^{T}(-\mbf k)$ by anti-symmetry)\footnote{For simplicity, we have written Eq.~\ref{eq:eta} for a $\mbf{Q}=0$ pair, but arbitrary $\mbf{Q}$ is possible within the QGN construction. }
\bea
\eta^\dagger = \sum_{\mbf k, nm} F_{nm}(\mbf k)\,\gamma^\dagger_{\mbf k, n}\gamma^\dagger_{-\mbf k, m} \ ,
\label{eq:eta}
\eea
which obeys the central commutation relation
\bea
\label{eq:Setacom}
[S_{\mbf{q},I}, \eta^\dagger] = 0 \ .
\eea
If this equation is satisfied, then there are zero-energy ground states at every even particle number
\bea
\ket{0_N} = \mathcal{N}_N^{-1/2} \eta^{\dagger N} \ket{0}, \quad N = 0, \dots, \Nf N_{\text{cell}}/2 \ .
\eea
The normalization constant $\mathcal{N}_N$ can be determined explicitly but will not be required here. The states $\ket{0_N}$ are ground states because $S_{\mbf q,I} \ket{0} = 0$ and 
\bea
 H \ket{0_N} = \sum_{\mbf q, IJ} V_{IJ}(\mbf q)\;  S^{\dag}_{\mbf q, I} \,  [S_{\mbf q,J} ,\eta^{\dagger N} ]\ket{0} = 0
\eea
and $H$ is positive semi-definite. A more familiar representation of the ground state manifold are the BCS states $e^{z \eta^\dag}\ket{0}$, which explicitly break $U(1)$ charge symmetry.

A nontrivial solution to \Eq{eq:Setacom} can be constructed when the ``quantum geometric nesting" condition
\bea
\label{eq:QGNcond}
\exists \mathcal{N} & \text{ s.t. } \sum_{\alpha \beta} \Pi_{\alpha'\beta';\alpha\beta}\mathcal{N}_{\alpha\beta} = 0 \ ,\\
\Pi_{\alpha'\beta';\alpha\beta} \equiv &\frac{1}{N_\text{cell}}\sum_{{\bf k}} \left[P_{\alpha'\alpha}({\bf k}) Q^*_{\beta \beta' } (-{\bf k})+ (P\leftrightarrow Q) \right]
\eea
is obeyed. A nontrivial solution to this equation requires a compatibility between the particle and hole Hilbert spaces at $\pm \mbf{k}$ in the flat band, in which case there is a nonzero nesting matrix obeying $\mathcal{N}^\dag \mathcal{N} = 1$ (we assume no degeneracy in the particle-particle nesting channel). Given $\mathcal{N}$, the pairing operator generating the ground states is
\bea
F_{nm}({\mbf k}) \equiv [U^\dag(\mbf{k}) \mathcal{N} U^*(-\mbf{k})]_{nm} \ .
\eea
Finally, using \Eq{eq:QGNcond}, we note that $F(\mbf k)F^\dagger(\mbf k) = \mathbbm{1}$. To find parent Hamiltonians for the ground state, we construct an infinite family of $S_{\mbf{q},I}$ operators commuting with $\eta^\dag$ can be systematically constructed (see App.~\ref{app:QGN} for a brief review). They take the form in \Eq{eq:Sq} where for onsite interactions, $S_{I}(\mbf k + \mbf q, \mbf k) = U^\dag(\mbf{k}+\mbf{q}) S_I U(\mbf{k})$ and $S_I$ is a matrix in the orbital basis. There are two important properties of the QGN models to point out. First, an assumption of the construction is the reality condition 
\bea
S^{\dag}_{\mbf q,I} = S_{-\mbf q,I}
\eea
which is equivalent to the condition $S_{I}(\mbf k + \mbf q, \mbf k) = S_{I}(\mbf k, \mbf k+\mbf{q})^\dag$ or $S_I^\dag = S_I$. This property implies that 
\bea
S_{\mbf{R},I} &= \sum_\mbf{q} e^{- i \mbf{q} \cdot \mbf{R}} S_{\mbf q,I} = S_{\mbf{R},I}^\dag
\eea
so that $H = \sum_{\mbf{R},\mbf{R}'} V_{IJ}(\mbf{R}-\mbf{R}) S_{\mbf{R} I} S_{\mbf{R}',J}$ is of a generalized spin-spin form. We emphasize that this property is not required for $H$ to be frustration-free, and that the $S_{\mbf{R},I}$'s do not in general commute. Second, using the reality condition and $[S_{\mbf{q},I}, \eta^\dagger] = 0$, it follows that $[H, \eta^\dagger] = 0$. This means $H$ has an enlarged charge-$SU(2)$ symmetry group generated by $\eta,\eta^\dagger, N$. 

Thus far, we have shown that $H$ is a frustration-free Hamiltonian whose ground states form high dimensional irrep of its charge su(2) algebra. BCS states within this manifold break $U(1)$ spontaneously, leading to Goldstone modes. This means that $H$ must realize type-II Goldstones with a dispersion $\omega(\mbf{q}) = O(q^2)$ and not type-I, where $\omega(\mbf{q}) = O(|q|)$~\cite{PhysRevX.4.031057,annurev:/content/journals/10.1146/annurev-conmatphys-031119-050644}. This is consistent with recent theorem~\cite{PhysRevB.110.195140} on frustration-free models, admits an anomalous critical field theory description~\cite{HanKivelson25}, and has been borne out by explicit calculations in a variety of flat-band ferromagnets, including exact Goldstone and magnon spectra in twisted bilayer graphene~\cite{TBGV,AlaviradSau2020,KallinHalperin1984} and related collective-mode calculations~\cite{WuDasSarmaQAHF2020,KhalafSoftModes2020}.

\subsection{Fock-space fragmentation}
\label{sec:models}


Here we briefly review the fock-space fragmentation structure for neutral excitations. Using the reality condition, we see that for any operator $\mathcal{O}$ acting on the ground states, the action of $H$ has a double-commutator structure:
\bea
\label{eq:bosonnumber}
 H \mathcal{O} \ket{0_N} &= \sum_{\mbf q, IJ} V_{IJ}(\mbf q)\;  S_{-\mbf q,I} \,  S_{\mbf q,J} \mathcal{O} \ket{0_N} \\
&= \sum_{\mbf q, IJ} V_{IJ}(\mbf q)\;  S_{-\mbf q,I} \, [ S_{\mbf q,J}, \mathcal{O}] \ket{0_N} \\
&= \sum_{\mbf q, IJ} V_{IJ}(\mbf q)\; [ S_{-\mbf q,I}, \, [ S_{\mbf q,J}, \mathcal{O}] ]\ket{0_N} \ . \\
\eea
An important consequence of this fact is that $H$ is block diagonal in the basis of states\footnote{To see that this spans the full Hilbert space, we observe that any excited state in the $N$ particle number sector can be written as $\mathcal{O} \ket{0_N}$ where $\mathcal{O}$ is a sum of products of operators in the form $(\gamma^\dag \dots \gamma^\dag) (\gamma \dots \gamma)$, in which each factor contains $N$ fermionic operators. Using fermionic anti-commutation relations, $(\gamma^\dag \dots \gamma^\dag) (\gamma \dots \gamma)$ can be written instead as a sum of terms in the form $\gamma^\dag \gamma \dots \gamma^\dag \gamma$ with $1,\dots,N$ factors $\gamma^\dag \gamma$. The action of H is block diagonal on each string length.} spanned by operator strings $\mathcal{O}_m$ in the form 
\bea
\mathcal{O}_m = \prod_{a=1}^m \gamma^\dag B^a \gamma
\eea
where $\gamma$ is a column vector whose entries are $\gamma_{\mbf{k},n}$. To prove this, observe that $S_{\mbf{q},I}$ is in the form $\gamma^\dag S \gamma$, and \bea
[\gamma^\dag S \gamma, \gamma^\dag B \gamma] = \gamma^\dag [S,B] \gamma
\eea
forms a closed algebra. This means
\bea
[S_{\mbf{q},I}, \mathcal{O}_m] = \sum_a \gamma^\dag B^1 \gamma \dots \gamma^\dag [S,B^a] \gamma \dots \gamma^\dag B^m \gamma 
\eea
and similarly $[S_{-\mbf{q},I},[S_{\mbf{q},J}, \mathcal{O}_m]]$ is also an operator in the same form as $\mathcal{O}_m$. Thus the action of $H$ on the space of $ \mathcal{O}_m$ operators is closed. As such, the exponentially large Hilbert space breaks into polynomially large blocks indexed by $m$. 

Hence, we have shown that all excitations obey a super-selection rule which we call ``boson number conservation" because $\gamma^\dag B \gamma$ is an elementary neutral bosonic excitation. In particular, the space of 1-boson excitations yields exact eigenstates determined by the 2-body eigenvalue equation 
\bea
\label{eq:collboson}
 H \gamma^\dag_{\mbf{p}+\mbf{k},m} \gamma_{\mbf{k},n} \ket{0_N}
= \!\!\sum_{\mbf{k}'m'n'}\mathcal{R}_{\mbf{k}mn}^{\mbf{k}'m'n'}(\mbf{p}) \gamma^\dag_{\mbf{p}+\mbf{k}',m'} \gamma_{\mbf{k}',n'} \ket{0_N}\ .
\eea
An explicit expression for the Hermitian matrix $R(\mbf{p})$ can be found in \App{app:exmatrices}. Since there are $N_f^2 N_{\text{cell}}$ excitations at each $\mbf{p}$, the 1-boson excitation can be determined with only linear scaling. Defining the eigen-decomposition
\bea
R_{\mbf{k}mn}^{\mbf{k}'m'n'}(\mbf{p}) = \sum_\mu E_\mu(\mbf{p}) \mathcal{V}^\mu_{\mbf{k}mn}(\mbf{p}) \mathcal{V}^{\mu*}_{\mbf{k}'m'n'}(\mbf{p})
\eea
for $\mu = 0, \dots, N_f^2 N_{\text{cell}}-1$, we obtain the exact (unnormalized) eigenstates
\bea
H b_{\mbf{p},\mu}\!\ket{0_N} &= E_\mu(\mbf{p}) b_{\mbf{p},\mu}\! \ket{0_N}\!, \\ b_{\mbf{p},\mu} &= \sum_{\mbf{k}mn}\mathcal{V}^{\mu *}_{\mbf{k}mn}(\mbf{p}) \gamma^\dag_{\mbf{p}+\mbf{k},m} \gamma_{\mbf{k},n}
\eea
which contain gapless Goldstone modes of the broken $U(1)$ symmetry. Finally, although it is not important for the following argument, we should point out that the $\mathcal{O}_m$ basis is in general over-complete and contains all the eigenstates of each $m' < m$ sector. For instance, the 1-boson state $\bar{N} \ket{0_N} \propto \ket{0_N}$ is also in the $m=0$ (ground state) sector. 

This hidden symmetry is the key property that enables us to solve the superfluid weight and dynamical correlation functions exactly in terms of the 1-boson sector for all $n$, as we now show. This allows for the analytical solution of various many-body response functions as we now show. 

\section{Superfluid Stiffness}
\label{sec:SFW}

In this section, we prove our main result. The key idea is to show that the many-body energy eigenvalue in the Kohn formula is equal, at leading order, to the expectation of a generalized BCS wavefunction thanks to boson conservation.  

\subsection{Vector Potential}

To compute the stiffness from Eq. \ref{eq:KohnDsexpression}, we need to thread a constant vector potential through the real-space torus defined by periodic boundary conditions $L_1 \mbf{a}_1 \times L_2 \mbf{a}_2$. In the unprojected model, this operation takes 
$c^\dag_{\mbf{R},\al} c_{\mbf{R}',\beta} \to e^{i \mbf{A} \cdot (\mbf{R} - \mbf{R}')} c^\dag_{\mbf{R},\al}c_{\mbf{R}',\beta} $.  In momentum space, this is the usual substitution $\mbf k \to \mbf k + \mbf A$ so that we obtain the perturbed Hamiltonian in the flat band Hilbert space:
\bea
 H(\mbf A) &= \sum_{\mbf q, IJ}V_{IJ}(\mbf q)\, S_{\mbf q,I}(\mbf A)^\dagger\, S_{\mbf q,J}(\mbf A),\\
 S_{\mbf q,I}(\mbf A) &= \frac{1}{\Vol}\sum_{\mbf k}\gamma^\dagger_{\mbf k+\mbf q}\,S_I(\mbf k + \mbf q + \mbf A,\, \mbf k + \mbf A)\,\gamma_{\mbf k}.
\label{eq:HAfam}
\eea
For brevity, we write $\del_i S_{\mbf q,J}(\mbf A)|_{\mbf{A}=0} = \del_i S_{\mbf q,J}$ henceforth. Note that $ H(\mbf A)$ remains a positive-semi-definite, so its ground energy satisfies $E(\mbf A) \ge 0$. Additionally, the spectrum is periodic in $\mbf{A}_i  = \mbf{A} \cdot \mbf{a}_i = \frac{2\pi}{L_i}$. We will always assume $\mbf{A}_i \ll 2\pi/L_i$ consistent with the order of limits in Eq. \ref{eq:KohnDsexpression}. 

Our first target is the superfluid weight
\bea
\label{eq:KohnDsexpression}
[D_s]_{ij} &= \lim_{L \to \infty} \lim_{\mbf{A} \to 0} \frac{1}{4\Vol}\,\del_{i}\del_{j}E(\mbf A)\ .
\eea
At zero temperature this curvature is the Drude weight; the pairing gap of the QGN models eliminates any normal-current response, so it coincides with the superfluid weight \cite{SWZ93,GaoHanKhalaf2026}.

\subsection{Kohn Formula}

We will calculate $\del_{i}\del_{j}E(\mbf A)$ by perturbation theory. To do so, we expand out the Hamiltonian in $A_i$: 
\be
 H(\mbf A) =  H + A_i H_i + \frac{1}{2} A_iA_j H_{ij} + O(A^3),
\ee
where the current operator in the projected Hilbert space is
\be
H_i = \sum_{\mbf q,IJ}V_{IJ}(\mbf q)(\del_i S_{-\mbf q,I}\, S_{\mbf q,J} +  S_{-\mbf q,I}\,\del_i S_{\mbf q,J}),
\label{eq:Hi}
\ee
and the susceptibility operator is
\bea
H_{ij}&= \sum_{\mbf q,IJ}V_{IJ}(\mbf q)\Big[ \del_i S_{-\mbf q,I}\, \del_j S_{\mbf q,J} +  \del_j S_{-\mbf q,I}\,\del_i S_{\mbf q,J} \\
&\quad +\del_{i} \del_{j} S_{-\mbf q,I}\, S_{\mbf q,J} +  S_{-\mbf q,I}\,\del_{i} \del_{j} S_{\mbf q,J}
\Big] \ .
\label{eq:Hij}
\eea
We easily see that the first order correction $\del_i E(\mbf{0}) = \braket{0_N|H_i|0_N}$ vanishes because $S_\mbf{q}^J\ket{0_N} = 0$ and $\bra{0_N} S_{-\mbf{q}}^I= 0$. The stiffness is then given by the second order formula
\bea
\del_{i}\del_{j}E(\mbf 0) = \braket{0_N|H_{ij}|0_N}
- 2\Re \!\sum_{E_M>0}\!\frac{\braket{0_N| H_i|M}\braket{M| H_j|0_N}}{E_M}
\label{eq:core}
\eea
where the sum is over all excited states of $H$. The key step to simplifying this sum is to observe that boson number conservation ensures that
\bea
\label{eq:currentoverlap}
\braket{M| H_j|0_N} &= \sum_{\mbf q,IJ}V_{IJ}(\mbf q)\bra{M}\del_i S_{-\mbf q,I}\, S_{\mbf q,J} +  S_{-\mbf q,I}\,\del_i S_{\mbf q,J} \ket{0_N} \\
&= \sum_{\mbf q,IJ}V_{IJ}(\mbf q)\bra{M}[\del_i S_{-\mbf q,I}, S_{\mbf q,J}] \ket{0_N} \\
\eea
is nonzero only if $\ket{M}$ is a 1-boson state because $[\del_i S_{-\mbf q,I}, S_{\mbf q,J}] \ket{0_N} = \sum_{\mu} c_\mu b_{\mbf{0},\mu} \ket{0_N}$ is supported only on the 1-boson states. At this point, Eq. \ref{eq:core} can be evaluated exactly using only the 1-boson excitation spectrum, since all expectation values $\braket{0_N| b^\dag_{\mbf{0},\mu} H_j|0_N}$ and $\braket{0_N|H_{ij}|0_N}$ can be evaluated exactly with Wick's theorem. However, there is a further simplification as we now show. 

\subsection{Trial Wavefunction}

Instead of evaluating the sum directly, we will identify a trial wavefunction whose variational energy is exactly $E(\mbf A)$ to $O(A^3)$, and thereby reproduces $\del_{i} \del_jE(\mbf{0})$. This wavefunction will be the optimized BCS-type wavefunction considered in~\cite{GaoHanKhalaf2026,Chiral}. This will show that the variational BCS space encompasses the exact wavefunction at leading order. 

Recall that the \emph{first} order perturbation theory expression for the eigenstate, 
 \bea
 \label{eq:1storderpsi}
\ket{0_N(\mbf{A})} &= \ket{0_N} - A_i \sum_{E_M>0} \frac{\braket{M| H_i|0_N}}{E_M} \ket{M} + O(A^2) \ ,
 \eea
yields an energy accurate to \emph{second} order:
\bea
\braket{0_N(\mbf{A})|H(\mbf{A})|0_N(\mbf{A})} = \frac{1}{2} A_iA_j \del_{ij} E(\mbf{0}) + O(A^3) \ .
\eea
Hence, if we can identify a trial state $\ket{\Psi(\mbf{A})}$ which reproduces the exact eigenstate to leading order, meaning $\ket{N(\mbf{A})} = \ket{\Psi(\mbf{A})} + O(A^2)$, then 
\bea
\braket{\Psi(\mbf{A})|H(\mbf{A})|\Psi(\mbf{A})}  = \frac{1}{2} A_iA_j \del_{ij} E(\mbf{0}) + O(A^3) \ ,
\eea 
In general, \emph{any} trial state $\ket{\Psi(\mbf{A})}$ produces an upper bound on $E(\mbf{A})$, and hence an upper bound on $\del_{i}\del_j E(\mbf{0}) \propto D_s$ since $E(\mbf{0}) = 0$. Indeed, in~\cite{GaoHanKhalaf2026,Chiral}, an upper bound on $\del_{i} \del_{j} E(\mbf{0})$ was constructing the projected-BCS state 
\bea
\ket{\Psi(A)} &\propto \left( \eta^{\dag}[F_\mbf{A}] \right)^N \ket{0}
\eea
where, defining a general BCS wavefunction, 
\bea
\label{eq:generalBCS}
\eta^\dagger[F_\mbf{A}] = \sum_{\mbf k, nm} [F_{\mbf{A}}(\mbf k)]_{mn}\,\gamma^\dagger_{\mbf k, n}\gamma^\dagger_{-\mbf k, m}
\eea
and varying $[F_\mbf{A}]$ to obtain the lowest-energy trial wavefunction. The result obtained from this trial wavefunction, for all QGN models, is
\bea
\label{eq:Dsupper}
[D_s]_{ij} &\leq \frac{N_f}{2} \nu (1-\nu) [m^{-1}_{\text{pair}}]_{ij} \ .
\eea
We now show that this trial wavefunction is exact to leading order, and hence \Eq{eq:Dsupper} is actually an equality.

First, we use the commutator $[\gamma_{\mbf{k},n} , \eta^\dag] = 2 \sum_m F_{n m}(\mbf{k})  \gamma^\dag_{-\mbf{k} m} $ to show the BdG identity
\bea
 \gamma_{\mbf{k},n} \eta^{\dag N} \ket{0} 
&=  2 N \sum_m F_{nm}(\mbf{k}) \gamma^\dag_{-\mbf{k},m} \, \eta^{\dag (N-1)} \ket{0} \ . \\
\eea
Multiplying both sides by $\gamma^\dag_{\mbf{k},m}$ yields
\bea
\label{eq:1pair}
\sum_{\mbf{k} mn} B_{mn}(\mbf{k}) \gamma^\dag_{\mbf{k} m}  \gamma_{\mbf{k},n} \eta^{\dag N} \ket{0} &= 2N \eta^\dag[BF] \,  \eta^{\dag (N-1)} \ket{0} 
\eea
or, in other words, that a single neutral excitation on the condensate deforms exactly one pair. To relate this to the variational states, we consider all variations of the BCS trial wavefunction $(\eta^{\dag}[F + \delta F])^N \ket{0}$ (see \Eq{eq:generalBCS}), which at $\delta F = 0$ is the exact ground state $\eta^{\dag N}\ket{0}$ for $\mbf{A}=0$. Differentiating, we obtain
\bea
\label{eq:derivative}
(\eta^{\dag}[F + \delta F])^N \ket{0} &= N \, \eta^\dag[\delta F] \, \eta^{\dag (N-1)}[F] \ket{0} + O(\delta F^2)  \\
\eea
because all pairs are commuting. Together, \Eq{eq:1pair} and \Eq{eq:derivative} show that to $O(\delta F^2)$, the BCS ansatz $(\eta^{\dag}[F + \delta F])^N \ket{0}$ spans all 1-boson excitations. Finally, since \Eq{eq:1storderpsi} shows that $\ket{N(\mbf{A})}$ is contained in the space of 1-boson excitations to $O(A^2)$, we see that there exists a BCS ansatz which is exactly $\ket{N(\mbf{A})}$ to leading order in $\mbf{A}$. Hence, as a variational wavefunction, it achieves the exact stiffness. 

\section{Dynamical Correlation Functions}
\label{sec:response}

The superfluid stiffness is a low-energy property, but in fact we can prove that dynamical correlation functions -- probing the excitations at finite $\omega$ or $\mbf{q}$ -- can also be determined exactly thanks to boson number conservation. We will illustrate this result with three examples: the optical conductivity, the dynamical structure factor, and electron Green's function. 

Recall that the formula for the optical conductivity is
\bea
\sigma_{ij}(\omega) &= \frac{1}{N_{\text{cell}}} \sum_M \frac{\braket{0_N|H_i|M} \braket{M|H_j|0_N}}{\omega} \delta(\omega - E_M)
\eea
since $H_i$ is the current, and the (projected) dynamical structure factor, for the flat band density operator $\bar{\rho}_{\mbf{q},\al}$, is 
\bea
{\cal S}_{\alpha \beta}(\mbf{q},\omega) &= \frac{1}{N_\text{cell}} \sum_{M} \braket{0_N|\bar{\rho}^\dag_{\mbf{q},\alpha}|M} \braket{M|\bar{\rho}_{\mbf{q},\beta}|0_N} \delta(\omega - E_M) 
\eea
Both of these quantities include matrix elements of 1-body operators, e.g. $\braket{M|\bar{\rho}_{\mbf{q} \al}|0_N}$ and $\braket{M|H_i|0_N}$. Due to boson number conservation, the sum over $M$ truncates to the 1-body excitations only. Hence, just like $D_s$, they can be evaluated using the solution to the 2-body problem. A detailed derivation is given in \App{app:optcond} and \App{app:projSq}. For the optical conductivity, we find the expression
\bea
\sigma_{ij}(\omega) &= 2 N_f \nu(1-\nu) \omega \sum_{\mu >0} \Tr \, \mathcal{P}_\mu \del_i \mathcal{P}_0 \del_j \mathcal{P}_0 \  \delta(\omega - E_\mu) \ . \\
\eea
where $\mathcal{P}_\mu$ is the projector onto the $\mu$th bosonic mode wavefunction of \Eq{eq:collboson} at $\mbf{p}=0$. Returning to the projected structure factor, we given an explicit expression in \App{app:projSq}, where we also prove the sum rule
\bea
\int {\cal S}_{\alpha \beta}(\mbf{q},\omega) d\omega &=  \frac{2 \nu(1-\nu)}{N_\text{cell}} \sum_{\mbf{k}}  P_{\alpha \beta}(\mbf{k}+\mbf{q}) P_{\beta \alpha}(\mbf{k}) \ . \\
\eea
The curvature of the largest eigenvalue of this matrix was shown in Ref. \cite{HA22} to be exactly the minimal quantum metric\cite{Huhtinen22, PhysRevB.102.165148,2026arXiv260721581W} when all orbitals are symmetry-related.

The last quantity we consider the the one-body Green's function in the Lehmann representation:
\bea
\label{eq:Gmn}
G_{mn}(\mbf{k},\omega) &= \sum_M \frac{\braket{0_N|\gamma_{\mbf{k} m}|M} \braket{M|\gamma^\dag_{\mbf{k} n}|0_N}}{\omega - E_M+ i 0^+} \\
&\quad +\sum_M \frac{\braket{0_N|\gamma^\dag_{\mbf{k} n}|M} \braket{M|\gamma_{\mbf{k} m}|0_N}}{\omega + E_M+ i 0^+} \ .
\eea
Again using \Eq{eq:bosonnumber}, the double commutator relation shows that $\gamma^\dag_{\mbf{k} n} \ket{0_N}$ are closed under $H$ and obey:
\bea
\label{eq:fermion}
 H \gamma^\dag_{\mbf{k},m} \ket{0_N} 
&= \sum_n \Sigma_{mn}(\mbf{k}) \gamma^\dag_{\mbf{k},n} \ket{0_N} \\
\Sigma_{mn}(\mbf{k}) &= \sum_{\mbf q, IJ} V_{IJ}(\mbf q) [S_I(\mbf{k},\mbf{k}+\mbf{q})S_J(\mbf{k}+\mbf{q},\mbf{k})]_{m'm} \ . \\
\eea
This means that all matrix elements in Eq. \ref{eq:Gmn} can be evaluated exactly from Wick's theorem, and hence the Green's function is also exactly calculable. Our result is 
\bea
G(\mbf{k},\omega) &= \frac{(1-\nu)}{\omega - \Sigma(\mbf{k})+ i 0^+}+  \frac{\nu}{\omega + \Sigma(\mbf{k}) + i 0^+} \ .
\eea
These examples, relevant to typically experimental probes of superconductivity, are illustrative of the general solvable structure of dynamical correlation functions. A $p$-particle and $h$-hole dynamical correlation function at any density can be computed exactly as a $p+h$-body problem. 



\section{Generalizations to spin systems}

The fragmentation mechanism is neither specific to QGN models nor restricted to fermionic Fock space. More generally, the double-commutator structure of Eq.~\ref{eq:double_comm_intro} can generate an exact grading of Krylov space by representations of an operator algebra. Here we illustrate this possibility with a simple spin model.

Consider spin-\(1\) local moments \(S_j^a\), and define
\begin{align}
    K_j^+ = \frac{1}{2}(S_j^+)^2,\qquad
    K_j^- = \frac{1}{2}(S_j^-)^2,\qquad
    K_j^z = \frac{1}{2}S_j^z .
\end{align}
These operators satisfy an \(SU(2)\) algebra. On each site, the spin-\(1\) Hilbert space decomposes under this algebra as $\mathbf{\frac12}\oplus\mathbf 0$
where \(\ket{+1}\) and \(\ket{-1}\) form the \(K=\frac12\) doublet, while \(\ket{0}\) is an \(SU(2)\) singlet. 

We now construct, on an arbitrary lattice and in arbitrary spatial dimension, the Hamiltonian
\begin{align}
    H
    =
    \sum_{ij,ab}
    V^{ab}_{ij}\,K_i^a K_j^b,
    \qquad
    V\succeq 0,
    \label{eq:spin_fragmentation_model}
\end{align}
where $[K_i^a,K_j^b] =i \delta_{ij} \epsilon^{abc} K^c_i$. Since the kernel \(V\) is positive semi-definite, \(H\) is positive semi-definite. An exact frustration-free ground state is
\begin{align}
    \ket{\mathrm{GS}}
    =
    \prod_j \ket{0}_j ,
\end{align}
which obeys
\begin{align}
    K_j^a\ket{\mathrm{GS}}
    =
    (K_j^a)^\dagger\ket{\mathrm{GS}}
    =0
\end{align}
for every site \(j\) and component \(a\). We note that Ref.~\cite{
PhysRevLett.133.176001} contains a similar construction corresponding to a special choice of $V^{ab}_{ij}$ in one dimension. Consequently, for any operator \(\mathcal O\),
\begin{align}
    [H,\mathcal O]\ket{\mathrm{GS}}
    &=
    \sum_{ij,ab}
    V^{ab}_{ij}
    \big[K_i^a,[K_j^b,\mathcal O]\big]\ket{\mathrm{GS}}
    \label{eq:spin_double_comm}
\end{align}
defines an \(SU(2)\)-invariant superoperator acting on operator space. Decomposing \(\mathcal O\) into irreducible tensor operators under the adjoint action of the global generators $K_{\rm tot}^a=\sum_j K_j^a$, the double-commutator in Eq.~\eqref{eq:spin_double_comm} cannot mix operators belonging to inequivalent \(SU(2)\) irreducible representations. Therefore, the Krylov spaces generated from \(\ket{\mathrm{GS}}\) are graded by the \(SU(2)\) representations. Concretely, we can generate all excitations from operator strings of the form $\ket{\pm_i}\bra{0_j}$, and the Hilbert space breaks into Krylov spaces corresponding to operator strings of fixed length.

This example provides a spin analogue of the Fock-space fragmentation: in fermionic QGN models, the invariant sectors are labeled by particle--hole composition \((p,h)\), whereas here they are labeled by irreducible representations of the spin \(SU(2)\). The construction of other models with solvable response functions according to this structure is left for future work.

\section{Discussions}

Several consequences follow immediately from our proof. First, for the attractive Hubbard model satisfying the uniform-pairing condition---the simplest member of the QGN family---the exact two-particle mass is determined by the minimal quantum metric~\cite{TormaLiangPeotta2018,Huhtinen22,HA22},
\begin{align}
[m_{\mathrm{pair}}^{-1}]_{ij}
=
\frac{2|U|}{N_{\mathrm{orb}}^\sigma}
[g_{\min}]_{ij},
\end{align}
where \(N_{\mathrm{orb}}^\sigma\) is the number of orbitals per spin. It follows that geometric, symmetry, and topological constraints that bound the integrated quantum metric from below~\cite{PT15,Xie20,HAPeri22} are promoted from statements about the two-particle problem or the mean-field superfluid weight to rigorous bounds on the true many-body stiffness.

Second, the argument is not specific to a vector potential. Consider a general perturbation \(h\). The essential requirement is that
\begin{align}
\partial_h \hat H\,\ket{0_N}
\end{align}
lies entirely within the single-particle--hole fragment of the condensate, which coincides with the tangent space of the BCS variational manifold. This condition is automatically satisfied when \(h\) enters smoothly through the fermion-bilinear building blocks of a QGN Hamiltonian: frustration freeness and Hermiticity reduce the action of \(\partial_h\hat H\) on the ground state to a fermion bilinear. It is also satisfied when \(\partial_h\hat H\) is itself a fermion bilinear, as for a field probing a continuous symmetry or the introduction of a weak single-particle dispersion. In either case, the exact first-order deformation of the ground state remains within the BCS tangent space, and the curvature \(d^2E/dh^2\) is therefore exactly equal to its BCS variational value. In particular, the bootstrap bounds on susceptibilities proposed in Ref.~\cite{GaoHanKhalaf2026} become equalities for this class of QGN models.

The Hermiticity of the QGN generators is crucial to this conclusion. In the frustration-free but non-QGN models of chiral flat-band superconductors in Ref.~\cite{Chiral}, the Hamiltonian is instead constructed from squares of \emph{non-Hermitian} bilinears whose Hermitian conjugates do not annihilate the ground state. The double-commutator reduction underlying Fock-space fragmentation then fails, and a perturbation can couple the ground state to sectors with higher particle--hole composition. The resulting current-carrying state therefore can acquire genuine many-body dressing beyond the BCS tangent space. Consistent with this distinction, the bootstrap lower bound on the stiffness in these models remains strictly below the BCS variational upper bound.

More abstractly, our result illustrates a broader connection between fragmentation and exact response. If a perturbation couples the ground state only to an invariant fragment of Hilbert space, and this fragment is fully contained in the tangent space of a variational manifold, then the first-order change of the wavefunction is captured exactly by that manifold and the associated second-order response is variationally exact.

{\bf Acknowledgment. } J.H.-A. gratefully acknowledges collaboration with Andrei Bernevig and P\"aivi T\"orm\"a on earlier related projects, and support from the Hertz Fellowship and the MIT Pappalardo Fellowship. Z.~H. is supported by the Gordon and Betty Moore Foundation EPiQS Award 8683. E.~K. is supported by NSF CAREER grant DMR award No. 2441781. We thank Zhengzhi Wu, Ming-Rui Li, and Hong Yao for coordinating the submission of an independent proof establishing the exactness of the superfluid stiffness formula~\cite{WuLiYao}. We used Claude Fable 5 and ChatGPT 5.6 Sol to help streamline the proof and polish the writing.

\newpage
\appendix
\onecolumngrid

\section{QGN models in a nutshell}

\label{app:QGN}

Here we give a brief review of the QGN construction. Various properties and examples may be found in \cite{QGN}.\\

A translation-invariant tight-binding model with $N_{\rm orb}$ orbitals (spin included unless otherwise specified) per unit cell is diagonalized by $\gamma_{\mbf k, n} = \sum_\al U^\dagger_{n\alpha}(\mbf k) c_{\mbf k, \alpha}$. We assume that bands $n = 1,\dots, \Nf$ are degenerate, exactly flat, and isolated by a gap $\Delta$ which we take to be infinitely large. The projector matrix $P_{\alpha\beta}(\mbf k) = \sum_{n\le\Nf}U_{\alpha n}(\mbf k)U^\dagger_{n\beta}(\mbf k)$ defines the Hilbert space of the flat band. Interactions are projected to the flat bands by restricting the band indices to $n \le \Nf$. Because we consider microscopic interactions in the form $c^\dag c c^\dag c$, projection induces one-body ``Hartree-Fock terms" $\bar{H}_{HF} = \bar{c}^\dag \braket{c c^\dag}_{\text{remote}} \bar{c}$, where $\bar{c}$ denotes projection to the flat bands. We will drop $\bar{H}_{HF}$ in this work. This can be justified in two ways. Firstly, we can require additional symmetries in the bands or interaction to ensure $\braket{c c^\dag}_{\text{remote}}  \propto \mathbbm{1}$ in which case $H_{HF}$ is a trivial chemical potential (see Ref. \cite{HA22} for an example with the uniform pairing condition.) Second, we can add a fine-tuned kinetic energy term to the Hamiltonian that exactly cancels the interaction-generated $H_{HF}$ term (see Ref. \cite{Chiral} for a physically motivated example).

Now we consider what properties of the bands are required to formulate a frustration-free model of superconductivity with Fock space fragmentation. Such a property is the ``QGN" condition. For simplicity, we restrict to the $\mbf Q = 0$ case where the condensing Cooper pair has zero total momentum. The general case may be found in \cite{QGN}. The QGN condition boils down to the simple assumption that there exists an antisymmetric $N_{\rm orb}\times N_{\rm orb}$ matrix $\mathcal N$ (the \emph{nesting matrix}) such that
\be
F_{nm}(\mbf k) \equiv \big[U^{\dag}(\mbf k)\, \mathcal N\, U^*(-\mbf k)\big]_{nm}
\label{eq:Ffull}
\ee
has no matrix elements between flat bands and remote bands for any $\mbf k$, ie that $F(\mbf{k})$ is block diagonal. The flat band block of $F(\mbf k)$ defines $\eta^\dagger$ in Eq.~\eqref{eq:eta}. Note that if such an $\mathcal{N}$ exists, then $\mathcal{N}\mathcal{N}^\dag$ commutes with $P(\mbf{k})$. Assuming $\mathcal{N}$ is non-degenerate (having no zero singular values), we can always take $\mathcal{N} \to (\mathcal{N}\mathcal{N}^\dag)^{-1/2}\mathcal{N}$ so that $\mathcal{N}$ can be assumed unitary. 

Given this assumption, we seek local hermitian bilinears $S_{\mbf R, I}$ obeying $[S_{\mbf R,I}, \eta^\dagger] = 0$. Writing the projected Fourier components as in Eq.~\eqref{eq:Sq}, with $[S_I(\mbf p, \mbf q)]_{mn} = [U^\dagger(\mbf p)S_{\rm orb, I}(\mbf p,\mbf q)U(\mbf q)]_{mn}$ with hermiticity enforcing $S_I(\mbf p, \mbf q) = S_I(\mbf q,\mbf p)^\dagger$, the commutation requirement is equivalent to
\be
F^{T}(-\mbf p)\, S^T_I(-\mbf q, -\mbf p) = S_I(\mbf p, \mbf q)\, F(\mbf q)
\quad \forall \mbf p, \mbf q. 
\label{eq:commcond}
\ee
This has infinitely many local solutions for every nesting matrix. For example, there is a family of solutions of the form:
\begin{align}\label{eq: separable form}
    S_{\text{orb},I}({\bf p},{\bf q}) = A_I({\bf p},{\bf q}) B^I_{\mu\nu} 
\end{align}
where $B$ is a hermitian matrix, and $A_I({\bf p},{\bf q})  = \left[A_I({\bf q},{\bf p})\right]^*$ for hermicity of $ S_{\mbf R,I}$. $A$ and $B$ are solutions of
\begin{align}
    A^{(I)}({\bf p},{\bf q}) = A^{(I)} (-{\bf q},-{\bf p}), \qquad \mathcal{N} B + B^T \mathcal{N} = 0 \ .
\end{align}
There are infinitely many solutions satisfying locality.

Since we have assumed the unitary case for $\mathcal{N}$ and $\mbf{Q}$ =0 pair momentum, the QGN relation simply shows that 
\bea
\mathcal{N}K\,P(\mbf k)K \mathcal{N}^\dag = P(-\mbf k)
\eea
where $K$ is complex conjugation. Hence the QGN condition is statement on the existence of an effective Kramers time-reversal symmetry. This is because $(\mathcal{N}K)^2 = \mathcal{N} \mathcal{N}^* = -\mathcal{N} \mathcal{N}^\dag = - \mathbbm{1}$. Similarly, it can be shown that the condition on $S^\dag_{\mbf{R},i}$ to commute with $\eta^\dag$ is that $S_{\text{orb},I}$ is odd under $\mathcal{N}K$.

\section{Collective Modes and Dynamical Correlation Functions}

In this section, we give explicit expressions for the elementary fermionic and bosonic excitation spectrum and eigenstates. We prove a handful of identities that simplify the evaluation of dynamical correlation functions.

\subsection{Collective Mode Hamiltonian and Identities}
\label{app:exmatrices}

First we derive the excitation spectrum for the elementary fermionic and bosonic modes. Using the basic commutators
\bea
[S_{\mbf q, I},\gamma_{\mbf{k},n}] &= -\frac{1}{\sqrt{\Vol}}\sum_{m}[S_I(\mbf k, \mbf k - \mbf{q})]_{nm}\gamma_{\mbf k - \mbf{q}, m} \ ,  \\
[S_{\mbf q, I},\gamma^\dag_{\mbf{k},n}] &= \frac{1}{\sqrt{\Vol}}\sum_{m}\,
\gamma^\dagger_{\mbf k + \mbf q, m}\, [S_I(\mbf k + \mbf q, \mbf k)]_{mn} \ , \\
\eea
we derive the excitation matrices
\bea
H \gamma^\dag_{\mbf{k},n} \ket{0_N} &=  \sum_{\mbf q, IJ} V_{IJ}(\mbf q) [ S_{-\mbf q,I},  [ S_{\mbf q, J},\gamma^\dag_{\mbf{k},n}]]\ket{0_N} \\
 &=  \sum_{\mbf q, IJ} V_{IJ}(\mbf q) [ S_{-\mbf q,I},  \frac{1}{\sqrt{\Vol}}\sum_{m}\,
\gamma^\dagger_{\mbf k + \mbf q, m}\, [S_J(\mbf k + \mbf q, \mbf k)]_{mn}]\ket{0_N} \\
&= \sum_{m} \gamma^\dagger_{\mbf k , m}\ket{0_N} \Sigma_{mn}(\mbf{k}), \qquad \Sigma_{mn}(\mbf{k}) = \frac{1}{\Vol} \sum_{\mbf q, IJ} V_{IJ}(\mbf q)  [S_I(\mbf k, \mbf k + \mbf q) S_J(\mbf k + \mbf q, \mbf k)]_{mn}  \\
H \gamma_{\mbf{k},n} \ket{0_N} &= \sum_{m} \Sigma_{nm}(\mbf{k}) \gamma_{\mbf k , m}\ket{0_N}
\eea
where $\Sigma_{mn}(\mbf{k})$ is the self-energy. Next we derive the bosonic excitation matrix
\bea
H \gamma^\dag_{\mbf{k}+\mbf{p},m}\gamma_{\mbf{k},n} \ket{0_N} &= \sum_{\mbf{q},IJ} V_{IJ}(\mbf{q}) S_{-\mbf{q},I}S_{\mbf{q},J}\gamma^\dag_{\mbf{k}+\mbf{p},m}\gamma_{\mbf{k},n} \ket{0_N}  \\
&= ([H, \gamma^\dag_{\mbf{k}+\mbf{p},m}]\gamma_{\mbf{k},n}+ \gamma^\dag_{\mbf{k}+\mbf{p},m}[H,\gamma_{\mbf{k},n}]) \ket{0_N}  \\
&= (\sum_{\mbf{q},IJ} V_{IJ}(\mbf{q}) [S_{-\mbf{q},I}S_{\mbf{q},J}, \gamma^\dag_{\mbf{k}+\mbf{p},m}]\gamma_{\mbf{k},n}+ \gamma^\dag_{\mbf{k}+\mbf{p},m}\Sigma_{nn'}(\mbf{k})\gamma_{\mbf{k},n'}) \ket{0_N}  \\
&= (\sum_{\mbf{q},IJ} V_{IJ}(\mbf{q}) (S_{-\mbf{q},I}[S_{\mbf{q},J}, \gamma^\dag_{\mbf{k}+\mbf{p},m}]+[S_{-\mbf{q},I}, \gamma^\dag_{\mbf{k}+\mbf{p},m}]S_{\mbf{q},J})\gamma_{\mbf{k},n}+ \gamma^\dag_{\mbf{k}+\mbf{p},m}\Sigma_{nn'}(\mbf{k})\gamma_{\mbf{k},n'}) \ket{0_N}  \\
&= (\sum_{\mbf{q},IJ} V_{IJ}(\mbf{q}) (S_{-\mbf{q},I}[S_{\mbf{q},J}, \gamma^\dag_{\mbf{k}+\mbf{p},m}]\gamma_{\mbf{k},n}+[S_{-\mbf{q},I}, \gamma^\dag_{\mbf{k}+\mbf{p},m}][S_{\mbf{q},J},\gamma_{\mbf{k},n}])+ \gamma^\dag_{\mbf{k}+\mbf{p},m}\Sigma_{nn'}(\mbf{k})\gamma_{\mbf{k},n'}) \ket{0_N}  \\
&= (\gamma^\dag_{\mbf{k}+\mbf{p},m}\Sigma_{nn'}(\mbf{k})\gamma_{\mbf{k},n'} + \gamma^\dag_{\mbf{k}+\mbf{p},m'} \Sigma_{m'm}(\mbf{k}+\mbf{p}) \gamma_{\mbf{k},n} + 2 \sum_{\mbf{q},IJ} V_{IJ}(\mbf{q}) [S_{\mbf{q},J}, \gamma^\dag_{\mbf{k}+\mbf{p},m}][S_{-\mbf{q},I},\gamma_{\mbf{k},n}]) \ket{0_N}  \\
&= \sum_{\mbf{k}'m'n'} \mathcal{R}_{\mbf{k}mn}^{\mbf{k}'m'n'}(\mbf{p}) \gamma^\dag_{\mbf{k}'+\mbf{p},m'}\gamma_{\mbf{k}',n'} \ket{0_N}
\eea
where the Hermitian matrix $\mathcal{R}(\mbf{p})$ is
\bea
\mathcal{R}_{\mbf{k}mn}^{\mbf{k}'m'n'}(\mbf{p}) &= \delta_{\mbf k,\mbf k'}(\delta_{nn'}\Sigma_{m'm}(\mbf k + \mbf{p}) + \delta_{mm'}\Sigma_{nn'}(\mbf k)) - \frac{2}{N_{\text{cell}}} \sum_{IJ} V_{IJ}(\mbf{k}'-\mbf{k}) [S_J(\mbf{k}'+\mbf{p},\mbf{k}+\mbf{p})]_{m' m}[S_I(\mbf{k},\mbf{k}')]_{nn'} \\
\eea
whose first term is the particle-hole continuum, and second term is the scattering matrix. The eigenvalue problem is $\mathcal{R}(\mbf{p}) \mathcal{V}^\mu(\mbf{p}) = E_\mu(\mbf{p}) \mathcal{V}^\mu(\mbf{p})$ corresponding to many-body eigenstates
\bea
b_{\mbf{p}\mu}\ket{0_N} &= \sum_{\mbf{k}mn}\mathcal{V}^{\mu *}_{\mbf{k}mn}(\mbf{p}) \gamma^\dag_{\mbf{k}+\mbf{p},m} \gamma_{\mbf{k},n}\ket{0_N} \ .
\eea
Note that $b_{\mu=0} = \sum_{\mbf{k}n}\gamma^\dag_{\mbf{k},n} \gamma_{\mbf{k},n} = \bar{N}$, the particle-number operator with $\mathcal{V}^{0*}_{\mbf{k}mn}(0) \propto \delta_{mn}$ and $E_0 = 0$. By orthogonality of the basis $\mathcal{V}^{\mu}(0)$, we have
\bea
\label{eq:orthotoN}
0 &= \sum_{\mbf{k} mn} \mathcal{V}^{\mu*}_{\mbf{k}mn}(0)\delta_{mn} = \sum_{\mbf{k}n} \mathcal{V}^{\mu*}_{\mbf{k}nn}(0), \qquad \mu > 0 \ . \\
\eea
We will use this property momentarily. 


We now prove an identity on the $b_{\mbf{p}, \mu}$ states. For convenience, we introduce matrix notation $\mathcal{V}^*_{\mbf{k}mn}(\mbf{p}) = [\mathcal{V}^*_{\mbf{k}}(\mbf{p})]_{mn}$. Then recall from the main text that
\bea
\sum_{\mbf{k} mn} \mathcal{V}^*_{\mbf{k}mn}(\mbf{p}) \gamma^\dag_{\mbf{k}+\mbf{p}, m}  \gamma_{\mbf{k},n} \eta^{\dag N} \ket{0} &= 2N \left( \sum_{\mbf{k}mn} [\mathcal{V}^*_{\mbf{k}}(\mbf{p}) F(\mbf{k})]_{mn}\gamma^\dag_{\mbf{k}+\mbf{p},m}\gamma^\dag_{-\mbf{k},n} \right) \,  \eta^{\dag (N-1)} \ket{0} \\
&= 2N \left( \sum_{\mbf{k}mn} [F(\mbf{k}+\mbf{p}) \mathcal{V}^\dag_{-\mbf{k}-\mbf{p}}(\mbf{p}) ]_{mn}\gamma^\dag_{\mbf{k}+\mbf{p},m}\gamma^\dag_{-\mbf{k},n} \right) \,  \eta^{\dag (N-1)} \ket{0} \\
&= \sum_{\mbf{k} mn} [F(\mbf{k}+\mbf{p})\mathcal{V}^\dag_{-\mbf{k}-\mbf{p}}(\mbf{p})F^\dag(\mbf{k})]_{mn} \gamma^\dag_{\mbf{k}+\mbf{p}, m}  \gamma_{\mbf{k},n} \eta^{\dag N} \ket{0} 
\eea
using anti-symmetry under the relabeling $\mbf{k} \to -\mbf{k}-\mbf{p}$, $m \leftrightarrow n$, which exchanges the two creation operators of the pair, together with $F^T(-\mbf{k}) = -F(\mbf{k})$. Hence, when diagonalizing $\mathcal{R}(\mbf{p})$, we will only keep a non-redundant basis of states where 
\bea
\label{eq:FVF}
\mathcal V_{\mbf kmn}(\mbf{p})=\sum_{m'n'}F^*_{n'm}(-\mbf k-\mbf{p})\,\mathcal V_{-\mbf k-\mbf{p},m'n'}(\mbf{p})\,F_{m'n}(-\mbf k)\ .
\eea
since the left-hand and right-hand sides create the same many-body eigenstate. Throughout this work, in all sums over $\mu$ it is implied that only the states obeying the anti-symmetry condition \Eq{eq:FVF} are kept. The other states with $\mathcal V_{\mbf kmn}(\mbf{p})=-\sum_{m'n'}F^*_{n'm}(-\mbf k-\mbf{p})\,\mathcal V_{-\mbf k-\mbf{p},m'n'}(\mbf{p})\,F_{m'n}(-\mbf k)$ are ``ghosts" which vanish for fermions (but are non-vanishing for bosons). 

Throughout the next calculations, we use the fact that expectation values of neutral operators $\mathcal{O}$ obey 
\bea
\lim_{N_{\text{cell}} \to \infty} \braket{0_N|\mathcal{O}|0_N} &= \braket{\mathcal{O}}
\eea
where $\braket{\cdot}$ is an expectation value in the BCS state with electron density $\nu = 2N/(N_{\text{cell}}N_f)$. 

\subsection{Optical Conductivity}
\label{app:optcond}

We start with the optical conductivity (we assume $\omega >0 $ in order to compute the regular part, since we have treated the Drude weight separately)
\bea
\label{eq:appsigma}
\sigma_{ij}(\omega) &= \frac{1}{N_{\text{cell}}} \sum_M \frac{\braket{0_N|H_i|M} \braket{M|H_j|0_N}}{\omega } \delta(\omega - E_M) \\
&= \frac{1}{N_{\text{cell}}} \sum_{\mu >0} \frac{\braket{H_i b_\mu}\braket{b^\dag_\mu H_j}}{\omega \braket{b_\mu^\dag b_\mu}} \delta(\omega - E_\mu) \\
&=  \frac{1}{N_{\text{cell}}}\sum_{\mu >0} \frac{\braket{H'_i b_\mu}\braket{b^\dag_\mu H'_j}}{\omega \braket{b_\mu^\dag b_\mu}} \delta(\omega - E_\mu) \\
\eea
where the relevant definitions are (note $ (H'_i)^\dag = - H'_i$)
\bea
H_i &= \sum_{\mbf q,IJ}V_{IJ}(\mbf q)(\del_i S_{-\mbf q,I}\, S_{\mbf q,J} +  S_{-\mbf q,I}\,\del_i S_{\mbf q,J}) \\
H'_i &= \sum_{\mbf q,IJ}V_{IJ}(\mbf q)[S_{-\mbf q,I},\del_i S_{\mbf q,J})] \\
&\equiv \sum_{\mbf{k} mn}[H'_i]_{\mbf{k}mn} \gamma^\dag_{\mbf{k}m} \gamma_{\mbf{k}n} \\
[H'_i]_{\mbf{k}mn} &=\frac{1}{\Vol} \sum_{\mbf{q} IJ} V_{IJ}(\mbf{q})
[\del_iS_I(\mbf k,\mbf k{+}\mbf q)\,S_J(\mbf k{+}\mbf q,\mbf k)
-S_J(\mbf k,\mbf k{-}\mbf q)\,\del_iS_I(\mbf k{-}\mbf q,\mbf k)]_{mn}
\eea
Next, we recall the elementary contractions
\bea
\braket{\gamma^\dag_{\mbf{k},m}\gamma_{\mbf{k},n}} = \nu \delta_{mn}, \ \braket{\gamma_{\mbf{k},m}\gamma_{-\mbf{k},n}} = -\sqrt{\nu(1-\nu)} F_{mn}(\mbf{k}), \ \braket{\gamma^\dag_{\mbf{k},n}\gamma^\dag_{-\mbf{k},m}} = \sqrt{\nu(1-\nu)} F^*_{nm}(\mbf{k}) \ .
\eea
Let us first compute the normalization factor in Eq. \ref{eq:appsigma}. 
\bea
\braket{b^\dag_{\mu}b_{\mu}} &= \sum_{\mbf{k}mn,\mbf{k}'m'n'}\mathcal{V}^{\mu*}_{\mbf{k}mn}(\mbf{0})\mathcal{V}^\mu_{\mbf{k}'m'n'}(\mbf{0}) \braket{\gamma_{\mbf{k}',n'}^\dag\gamma_{\mbf{k}',m'} \gamma^\dag_{\mbf{k},m} \gamma_{\mbf{k},n}} \\
&= \sum_{\mbf{k}mn,\mbf{k}'m'n'}\mathcal{V}^{\mu *}_{\mbf{k}mn}(\mbf{0})\mathcal{V}^\mu_{\mbf{k}'m'n'}(\mbf{0}) (\nu^2 \delta_{m'n'}\delta_{mn} + \nu(1-\nu) \delta_{\mbf{k},\mbf{k}'} \delta_{n'n}\delta_{m'm} + \nu(1-\nu) \delta_{\mbf{k},-\mbf{k}'} F_{m' n}(\mbf{k}') F^*_{n' m}(\mbf{k}') ) \\
&= \nu(1-\nu) \sum_{\mbf{k}mn}( \mathcal{V}^{\mu*}_{\mbf{k}mn}(\mbf{0})\mathcal{V}^\mu_{\mbf{k}mn}(\mbf{0})  +\sum_{m'n'}\mathcal{V}^{\mu*}_{\mbf{k}mn}(\mbf{0})\mathcal{V}^\mu_{-\mbf{k},m'n'}(\mbf{0}) F_{m' n}(-\mbf{k}) F^*_{n' m}(-\mbf{k}) ) \\
&= \nu(1-\nu) \left( 1 + \sum_{\mbf{k}mn,m'n'}\mathcal{V}^{\mu*}_{\mbf{k}mn}(\mbf{0})\mathcal{V}^\mu_{-\mbf{k},m'n'}(\mbf{0}) F_{m' n}(-\mbf{k}) F^*_{n' m}(-\mbf{k}) \right) \\
&= 2\nu(1-\nu) \\
\eea
where we dropped the Hartree term using Eq. \ref{eq:orthotoN} and used the identity in Eq. \ref{eq:FVF}. 

Next, we come to the numerator. The first fact we need is that the Hartree contraction $\braket{b^\dag_\mu}\braket{H'_i}$ in $\braket{b^\dag_\mu H'_i}$ vanishes for $\mu > 0$. This is because $\braket{b^\dag_\mu} \propto \sum_{\mbf{k} n}V^{\mu *}_{\mbf{k} nn} = 0$. 
Thus we have
\bea
\braket{b^\dag_\mu H'_i} &= \nu(1-\nu) \sum_{\mbf{k}mn,\mbf{k}'m'n'}[H'_i]_{\mbf{k}mn}\mathcal{V}^\mu_{\mbf{k}'m'n'}(\mbf{0}) (\delta_{\mbf{k},\mbf{k}'} \delta_{n'n}\delta_{m'm} + \delta_{\mbf{k},-\mbf{k}'} F_{m' n}(\mbf{k}') F^*_{n' m}(\mbf{k}') ) \\
&= \nu(1-\nu) \sum_{\mbf{k}mn} ([H'_i]_{\mbf{k}mn}\mathcal{V}^\mu_{\mbf{k}mn}(\mbf{0})  + \sum_{m'n'} [H_i]_{\mbf{k}mn}\mathcal{V}^\mu_{-\mbf{k},m'n'}(\mbf{0})  F_{m' n}(\mbf{k}') F^*_{n' m}(\mbf{k}') ) \\
&= 2\nu(1-\nu) \sum_{\mbf{k}mn} [H'_i]_{\mbf{k}mn}\mathcal{V}^\mu_{\mbf{k}mn}(\mbf{0})   \\
\eea
In fact, this expression can be simplified further using identities between $\Sigma$ and $H_i$. To see this, note the decomposition
\bea
T_i(\mbf k) &= \frac{1}{\Vol}\sum_{\mbf q,IJ}V_{IJ}(\mbf q)\,
\del_i S_J(\mbf k,\mbf k-\mbf q)\,S_I(\mbf k-\mbf q,\mbf k) \\
\del_i \Sigma(\mbf{k}) &= T_i^\dag(\mbf{k}) + T_i(\mbf{k})  \\
H'_i(\mbf{k}) &= T_i^\dag(\mbf{k}) - T_i(\mbf{k}) \ .
\eea
We now relate this quantity to $\mathcal{R}(\mbf{p})$. Observe that
\bea
\del_{p_i} \mathcal{R}_{\mbf{k}mn}^{\mbf{k}'m'n'}(\mbf{p}) &= \delta_{\mbf k,\mbf k'}\delta_{nn'}\del_{p_i}\Sigma_{m'm}(\mbf k + \mbf{p}) - \frac{2}{N_{\text{cell}}} \sum_{IJ} V_{IJ}(\mbf{k}'-\mbf{k}) [\del_{p_i} S_J(\mbf{k}'+\mbf{p},\mbf{k}+\mbf{p})]_{m' m}[S_I(\mbf{k},\mbf{k}')]_{nn'} \ .
\eea
Now summing over the Goldstone mode $[\mathcal{V}^{\mu=0}(\mbf{p}=0)]_{\mbf{k}mn} \propto \delta_{mn}$ gives
\bea
\sum_{\mbf{k} mn} \delta_{mn} \del_{p_i} \mathcal{R}_{\mbf{k}mn}^{\mbf{k}'m'n'}(\mbf{0}) &= \del_{i}\Sigma_{m'n'}(\mbf k') - \frac{2}{N_{\text{cell}}} \sum_{\mbf{k}}\sum_{IJ} V_{IJ}(\mbf{k}'-\mbf{k}) [\del_{i} S_J(\mbf{k}',\mbf{k})S_I(\mbf{k},\mbf{k}')]_{m'n'} \\
&= \del_{i}\Sigma_{m'n'}(\mbf k') - \frac{2}{N_{\text{cell}}} \sum_{\mbf{q}}\sum_{IJ} V_{IJ}(\mbf{q}) [\del_{i} S_J(\mbf{k}',\mbf{k}'-\mbf{q})S_I(\mbf{k}'-\mbf{q},\mbf{k}')]_{m'n'} \\
&= T^\dag_i(\mbf{k}) + T_i(\mbf{k}) - 2 T_i(\mbf{k}) \\
&= [H'_i(\mbf{k}')]_{m'n'} \ .
\eea
In matrix notation, this is
\bea
\sqrt{N_{\text{cell}} N_f} \, \mathcal{V}^{\dag}_0 \del_i \mathcal{R}(0) =  H'_i \ .
\eea
Differentiating the eigenvalue equation gives
\bea
 \mathcal{R}\del_i \mathcal{V}_\mu +  \del_i \mathcal{R} \, \mathcal{V}_\mu &= \del_i E \mathcal{V}_\mu + E \del_i \mathcal{V}_\mu \\
 \mathcal{V}^\dag_0\mathcal{R}\del_i \mathcal{V}_\mu +  \mathcal{V}^\dag_0 \del_i \mathcal{R} \, \mathcal{V}_\mu &= \del_i E \mathcal{V}^\dag_0 \mathcal{V}_\mu + E \mathcal{V}^\dag_0\del_i \mathcal{V}_\mu \\
 0 +  H_i \mathcal{V}_\mu/\sqrt{N_{\text{cell}} N_f} &= 0 + E \mathcal{V}^\dag_0\del_i \mathcal{V}_\mu \\
\eea
where the first zero is the fact that the Goldstone is at $E = 0$ and the second is orthogonality. Written out in indices, we find
\bea
\sum_{\mbf k mn}[H'_i]_{\mbf k mn}\,\mathcal{V}^\mu_{\mbf k mn}(\mbf 0)
= E_\mu(\mbf 0)\ \del_{p_i} \sum_{\mbf k,mn}\mathcal{V}^\mu_{\mbf{k}mn}(\mbf 0) \delta_{mn} \ .\\
\eea
Returning all the way to Eq. \ref{eq:appsigma}, we obtain
\bea
\sigma_{ij}(\omega) &= 2 \nu(1-\nu)  \frac{1}{N_{\text{cell}}} \sum_\mu \frac{1}{\omega} (\sqrt{N_{\text{cell}} N_f})^2 E_\mu^2 (\del_i \mathcal{V}_\mu^\dag) \mathcal{V}_0 \mathcal{V}_0^\dag (\del_j \mathcal{V}_\mu) \delta(\omega - E_\mu) \\
&= 2 N_f \nu(1-\nu) \omega \sum_\mu (\del_i \mathcal{V}_\mu)^\dag \mathcal{V}_0 \mathcal{V}_0^\dag (\del_j \mathcal{V}_\mu) \delta(\omega - E_\mu) \ . \\
\eea
To write this in a manifestly gauge-invariant way, we define projectors $\mathcal{P}_\mu = \mathcal{V}_\mu \mathcal{V}_\mu^\dag$. Then observe
\bea
(\del_i \mathcal{V}_\mu)^\dag \mathcal{V}_0 \mathcal{V}_0^\dag (\del_j \mathcal{V}_\mu) &= (\del_i \mathcal{V}_\mu)^\dag \mathcal{P}_0 \mathcal{P}_0 (\del_j \mathcal{V}_\mu) \\
&= \mathcal{V}_\mu^\dag \del_i \mathcal{P}_0 \del_j \mathcal{P}_0 \mathcal{V}_\mu \\
&= \Tr \, \mathcal{P}_\mu \del_i \mathcal{P}_0 \del_j \mathcal{P}_0 \ . \\
\eea
Hence we obtain our final form
\bea
\sigma_{ij}(\omega) &= 2 N_f \nu(1-\nu) \omega \sum_{\mu >0} \Tr \, \mathcal{P}_\mu \del_i \mathcal{P}_0 \del_j \mathcal{P}_0 \  \delta(\omega - E_\mu) \ . \\
\eea
This leads immediately to a geometric sum rule
\bea
\int \frac{\sigma_{ij}(\omega)}{\omega} d\omega &= 2 N_f \nu(1-\nu) \sum_{\mu > 0} \Tr \, \mathcal{P}_\mu \del_i \mathcal{P}_0 \del_j \mathcal{P}_0 \\
&= 2 N_f \nu(1-\nu) \Tr \, \del_j \mathcal{P}_0 (1- \mathcal{P}_0) \del_i \mathcal{P}_0  \\
\eea
which is the quantum metric of the collective modes at $\mbf{p}=0$. 

\subsection{Projected Structure Factor}
\label{app:projSq}

Next we want to compute the orbital-resolved projected structure factor
\bea
{\cal S}_{\alpha \beta}(\mbf{q},\omega) &= \frac{1}{N_\text{cell}} \sum_{M} \braket{0_N|\bar{\rho}^\dag_{\mbf{q},\alpha}|M} \braket{M|\bar{\rho}_{\mbf{q},\beta}|0_N} \delta(\omega - E_M) 
\eea
where the density operator is
\bea
\bar{\rho}_{\mbf{q},\al} &= \sum_\mbf{k} M^\al_{mn}(\mbf{k},\mbf{q})\gamma^\dag_{\mbf{k}+\mbf{q},m}\gamma_{\mbf{k},n}, \qquad M_{mn}(\mbf{k},\mbf{q}) = U^*_{\al m}(\mbf{k}+\mbf{q})U_{\al n}(\mbf{k}) \ .
\eea
Using boson conservation, we have
\bea
{\cal S}_{\alpha \beta}(\mbf{q},\omega) &= \frac{1}{N_\text{cell}} \sum_{\mu} \frac{\braket{b^\dag_{\mbf{q},\mu}\bar{\rho}_{\mbf{q},\alpha}}^\dag\braket{b^\dag_{\mbf{q},\mu}\bar{\rho}_{\mbf{q},\beta}}}{\braket{b^\dag_{\mbf{q} \mu} b_{\mbf{q} \mu}}} \delta(\omega - E_\mu(\mbf{q})) \ . 
\eea
First we compute the denominator. Since $\braket{b^\dag_{\mbf{q} \mu}} = 0$ for all $\mu > 0$, we obtain
\bea
\braket{b^\dag_{\mbf{q} \mu} b_{\mbf{q} \mu}} &= \sum_{\mbf{k}mn,\mbf{k}'m'n'} \mathcal{V}^{\mu }_{\mbf{k}'m'n'}(\mbf{q}) \mathcal{V}^{\mu *}_{\mbf{k}mn}(\mbf{q})\braket{\gamma_{\mbf{k}',n'}^\dag\gamma_{\mbf{k}'+\mbf{q},m'} \gamma^\dag_{\mbf{k}+\mbf{q},m} \gamma_{\mbf{k},n}} \\
&= \nu(1-\nu) \sum_{\mbf{k}mn,\mbf{k}'m'n'} \mathcal{V}^{\mu }_{\mbf{k}'m'n'}(\mbf{q}) \mathcal{V}^{\mu *}_{\mbf{k}mn}(\mbf{q})(\delta_{\mbf{k},\mbf{k}'} \delta_{nn'} \delta_{mm'} + \delta_{\mbf{k},-\mbf{k}'-\mbf{q}} F_{m' n}(\mbf{k}'+\mbf{q}) F^*_{n' m}(\mbf{k}')) \\
&= 2 \nu(1-\nu) \\
\eea
where we again restricted to the non-redundant  $\mathcal{V}^{\mu }_{\mbf{k}'m'n'}(\mbf{q})$ states. Moving to the numerator, the Hartree contraction $\braket{b^\dag_{\mbf{q},\mu}}\braket{\bar{\rho}_{\mbf{q},\al}}$ vanishes as before so that
\bea
\braket{b^\dag_{\mbf{q} \mu} \rho_{\mbf{q} \al}} &= \sum_{\mbf{k}mn,\mbf{k}'m'n'} \mathcal{V}^{\mu }_{\mbf{k}'m'n'}(\mbf{q}) M^\al_{mn}(\mbf{k},\mbf{q}) \braket{\gamma_{\mbf{k}',n'}^\dag\gamma_{\mbf{k}'+\mbf{q},m'} \gamma^\dag_{\mbf{k}+\mbf{q},m} \gamma_{\mbf{k},n}} \\
&= 2 \nu(1-\nu) \sum_{\mbf{k}mn} \mathcal{V}^{\mu }_{\mbf{k}mn}(\mbf{q}) M^\al_{mn}(\mbf{k},\mbf{q}) \ . \\
\eea
Thus we obtain the expression
\bea
{\cal S}_{\alpha \beta}(\mbf{q},\omega) &= \frac{2 \nu(1-\nu)}{N_\text{cell}} \sum_{\mu} \sum_{\mbf{k}mn,\mbf{k}'m'n'} (\mathcal{V}^{\mu }_{\mbf{k}'m'n'}(\mbf{q}) M^\alpha_{m'n'}(\mbf{k}',\mbf{q}'))^*\mathcal{V}^{\mu }_{\mbf{k}mn}(\mbf{q}) M^\beta_{mn}(\mbf{k},\mbf{q}) \delta(\omega - E_\mu(\mbf{q})) \ .
\eea
There is hence the sum rule
\bea
\int {\cal S}_{\alpha \beta}(\mbf{q},\omega) d\omega &=  \frac{2 \nu(1-\nu)}{N_\text{cell}} \sum_{\mu} \sum_{\mbf{k}mn,\mbf{k}'m'n'} (\mathcal{V}^{\mu }_{\mbf{k}'m'n'}(\mbf{q}) M^\alpha_{m'n'}(\mbf{k}',\mbf{q}))^*\mathcal{V}^{\mu }_{\mbf{k}mn}(\mbf{q}) M^\beta_{mn}(\mbf{k},\mbf{q}) \\
&=  \frac{2 \nu(1-\nu)}{N_\text{cell}} \sum_{\mbf{k}mn} M^\alpha_{mn}(\mbf{k},\mbf{q})^* M^\beta_{mn}(\mbf{k},\mbf{q}) \\
&=  \frac{2 \nu(1-\nu)}{N_\text{cell}} \sum_{\mbf{k}mn} M^\alpha_{mn}(\mbf{k},\mbf{q})^* M^\beta_{mn}(\mbf{k},\mbf{q}) \\
&=  \frac{2 \nu(1-\nu)}{N_\text{cell}} \sum_{\mbf{k}}  P_{\alpha \beta}(\mbf{k}+\mbf{q}) P_{\beta \alpha}(\mbf{k}) \\
\eea
where we have assumed $\mbf{q} \neq 0$ to use the completeness relation of the $\mathcal{V}^\mu$ eigenvectors since the $\mu$ sum excludes $\mu=0$ at $\mbf{q}=0$. We have also used the fact that $\bar{\rho}_{\mbf{q}}$ is properly anti-symmetrized as in \Eq{eq:FVF} so that the sum over $\mathcal{V}$ acts as the identity. 

The properties of the object
\bea
h_{\alpha \beta}(\mbf{q}) = \sum_{\mbf{k}}  P_{\alpha \beta}(\mbf{k}+\mbf{q}) P_{\beta \alpha}(\mbf{k})
\eea
were studied in Ref. \cite{HA22} in the case where all orbitals are symmetry-related yielding the so-called ``uniform pairing condition." In this case, it was shown that the largest eigenvalue of $h_{\alpha \beta}(\mbf{q})$ takes the form
\bea
\lambda_{\max}[h(\mbf{q})] &= \left( \frac{N_f}{N_{\text{orb}}}  - \frac{1}{2} \frac{q_i q_j}{N_{\text{orb}}} \int \frac{d^2k}{\Omega_{BZ}} g^{min}_{ij}(\mbf{k}) + \dots \right)
\eea
where $g^{min}$ is the minimal quantum metric. Hence we find the sum rule
\bea
\lim_{\mbf{q}\to 0} \del_i \del_j \max  \int {\cal S}(\mbf{q},\omega) d\omega &=  - \nu(1-\nu) \frac{g^{min}_{ij}}{N_{\text{orb}}}
\eea
Note that the coefficient of the $O(q^2)$ term is negative, which is characteristic of the \emph{projected} structure factor. 

\subsection{Electron Green's Function}

Let us first recall the self-energy matrix for the charge $\pm1$ excitations
\bea
\Sigma_{mn}(\mbf{k}) = \frac{1}{\Vol} \sum_{\mbf q, IJ} V_{IJ}(\mbf q)  [S_I(\mbf k, \mbf k + \mbf q) S_J(\mbf k + \mbf q, \mbf k)]_{mn} \ .
\eea
We will denote the eigenbasis by $\Sigma(\mbf{k}) = \sum_\mu \mathcal{U}_\mu(\mbf{k})\mathcal{U}^\dag_\mu(\mbf{k}) \epsilon_\mu(\mbf{k})$ with
\bea
\gamma^\dag_{\mbf{k} \mu} &= \sum_m \gamma^\dag_{\mbf{k},m} \mathcal{U}_{m \mu}(\mbf{k}), \qquad H \gamma^\dag_{\mbf{k} \mu} \ket{0_N} = \epsilon_\mu(\mbf{k})  \gamma^\dag_{\mbf{k} \mu} \ket{0_N} \ .
\eea
Now we consider the fermion Green's function
\bea
G_{mn}(\mbf{k},\omega) &= \sum_M \frac{\braket{0_N|\gamma_{\mbf{k} m}|M} \braket{M|\gamma^\dag_{\mbf{k} n}|0_N}}{\omega - E_M+ i 0^+}+\sum_M \frac{\braket{0_N|\gamma^\dag_{\mbf{k} n}|M} \braket{M|\gamma_{\mbf{k} m}|0_N}}{\omega + E_M+ i 0^+} \\
&= \sum_\mu \frac{1}{\braket{\gamma_{\mbf{k} \mu}\gamma^\dag_{\mbf{k} \mu}}} \frac{\braket{\gamma_{\mbf{k} m} 
\gamma^\dag_{\mbf{k} \mu}} \braket{\gamma_{\mbf{k} \mu}\gamma^\dag_{\mbf{k} n}}}{\omega - \epsilon_{\mu}(\mbf{k})+ i 0^+}+\sum_\mu \frac{1}{\braket{\gamma^\dag_{\mbf{k} \mu}\gamma_{\mbf{k} \mu}}} \frac{\braket{\gamma^\dag_{\mbf{k} n} \gamma_{\mbf{k}\mu}} \braket{\gamma_{\mbf{k}\mu}^\dag \gamma_{\mbf{k} m}}}{\omega + \epsilon_{\mu}(\mbf{k}) + i 0^+} \\
&= (1-\nu) \sum_\mu \frac{[\mathcal{U}_\mu(\mbf{k})\mathcal{U}_\mu^\dag(\mbf{k})]_{mn}}{\omega - \epsilon_{\mu}(\mbf{k})+ i 0^+}+ \nu \sum_\mu \frac{[\mathcal{U}_\mu(\mbf{k})\mathcal{U}_\mu^\dag(\mbf{k})]_{mn}}{\omega +\epsilon_{\mu}(\mbf{k})+ i 0^+} \\
&= \left[ \frac{(1-\nu)}{\omega - \Sigma(\mbf{k})+ i 0^+}+  \frac{\nu}{\omega + \Sigma(\mbf{k}) + i 0^+} \right]_{mn}\\
\eea
where we used the fact that the charge $\pm 1$ excitations eigenvalues are identical for particles and holes.

\bibliography{qgn_stiffness_refs}

@ARTICLE{2026arXiv260301922K,
       author = {{Kitamura}, Taisei and {Nakai}, Hiroki and {Katsura}, Hosho and {Arita}, Ryotaro},
        title = "{Quantum-geometry-driven exact ferromagnetic ground state in a nearly flat band}",
      journal = {arXiv e-prints},
         year = 2026,
        month = mar,
          eid = {arXiv:2603.01922},
        pages = {arXiv:2603.01922},
          doi = {10.48550/arXiv.2603.01922},
archivePrefix = {arXiv},
       eprint = {2603.01922},
 primaryClass = {cond-mat.str-el},
       adsurl = {https://ui.adsabs.harvard.edu/abs/2026arXiv260301922K}
}

@ARTICLE{2024arXiv240207171K,
       author = {{Kang}, Junha and {Oh}, Taekoo and {Lee}, Junhyun and {Yang}, Bohm-Jung},
        title = "{Quantum geometric bound for saturated ferromagnetism}",
      journal = {arXiv e-prints},
         year = 2024,
        month = feb,
          eid = {arXiv:2402.07171},
        pages = {arXiv:2402.07171},
          doi = {10.48550/arXiv.2402.07171},
archivePrefix = {arXiv},
       eprint = {2402.07171},
 primaryClass = {cond-mat.str-el},
       adsurl = {https://ui.adsabs.harvard.edu/abs/2024arXiv240207171K}
}

@article{PhysRevB.102.165148,
  title = {Contrasting lattice geometry dependent versus independent quantities: Ramifications for Berry curvature, energy gaps, and dynamics},
  author = {Simon, Steven H. and Rudner, Mark S.},
  journal = {Phys. Rev. B},
  volume = {102},
  issue = {16},
  pages = {165148},
  numpages = {13},
  year = {2020},
  month = {Oct},
  publisher = {American Physical Society},
  doi = {10.1103/PhysRevB.102.165148},
  url = {https://link.aps.org/doi/10.1103/PhysRevB.102.165148}
}

@ARTICLE{2026arXiv260721581W,
       author = {{Wang}, Ziwei and {Raca}, Charlie and {Simon}, Steven H.},
        title = "{Symmetry and Quantum Geometry in Bloch Bands}",
      journal = {arXiv e-prints},
         year = 2026,
        month = jul,
          eid = {arXiv:2607.21581},
        pages = {arXiv:2607.21581},
          doi = {10.48550/arXiv.2607.21581},
archivePrefix = {arXiv},
       eprint = {2607.21581},
 primaryClass = {cond-mat.str-el},
       adsurl = {https://ui.adsabs.harvard.edu/abs/2026arXiv260721581W}
}

@ARTICLE{2024PhRvL.132b6002C,
       author = {{Chen}, Shuai A. and {Law}, K.~T.},
        title = "{Ginzburg-Landau Theory of Flat-Band Superconductors with Quantum Metric}",
      journal = {\prl},
         year = 2024,
        month = jan,
       volume = {132},
       number = {2},
          eid = {026002},
        pages = {026002},
          doi = {10.1103/PhysRevLett.132.026002},
archivePrefix = {arXiv},
       eprint = {2303.15504},
 primaryClass = {cond-mat.supr-con},
       adsurl = {https://ui.adsabs.harvard.edu/abs/2024PhRvL.132b6002C}
}

@article{Kitaev2006,
  author  = {Kitaev, Alexei},
  title   = {Anyons in an Exactly Solved Model and Beyond},
  journal = {Annals of Physics},
  volume  = {321},
  number  = {1},
  pages   = {2--111},
  year    = {2006},
  doi     = {10.1016/j.aop.2005.10.005}
}

@article{LevinWen2005,
  author  = {Levin, Michael A. and Wen, Xiao-Gang},
  title   = {String-Net Condensation: A Physical Mechanism for Topological Phases},
  journal = {Physical Review B},
  volume  = {71},
  pages   = {045110},
  year    = {2005},
  doi     = {10.1103/PhysRevB.71.045110}
}

@article{Bethe1931,
  author  = {Bethe, H. A.},
  title   = {Zur Theorie der Metalle. I. Eigenwerte und Eigenfunktionen der linearen Atomkette},
  journal = {Zeitschrift f{\"u}r Physik},
  volume  = {71},
  pages   = {205--226},
  year    = {1931},
  doi     = {10.1007/BF01341708}
}

@article{Haldane1988,
  author  = {Haldane, F. D. M.},
  title   = {Exact Jastrow-Gutzwiller Resonating-Valence-Bond Ground State of the Spin-1/2 Antiferromagnetic Heisenberg Chain with $1/r^2$ Exchange},
  journal = {Physical Review Letters},
  volume  = {60},
  pages   = {635--638},
  year    = {1988},
  doi     = {10.1103/PhysRevLett.60.635}
}

@article{Shastry1988,
  author  = {Shastry, B. Sriram},
  title   = {Exact Solution of an $S=1/2$ Heisenberg Antiferromagnetic Chain with Long-Ranged Interactions},
  journal = {Physical Review Letters},
  volume  = {60},
  pages   = {639--642},
  year    = {1988},
  doi     = {10.1103/PhysRevLett.60.639}
}

@article{Calogero1971,
  author  = {Calogero, Francesco},
  title   = {Solution of the One-Dimensional $N$-Body Problems with Quadratic and/or Inversely Quadratic Pair Potentials},
  journal = {Journal of Mathematical Physics},
  volume  = {12},
  pages   = {419--436},
  year    = {1971},
  doi     = {10.1063/1.1665604}
}

@article{Sutherland1971,
  author  = {Sutherland, Bill},
  title   = {Exact Results for a Quantum Many-Body Problem in One Dimension},
  journal = {Physical Review A},
  volume  = {4},
  pages   = {2019--2021},
  year    = {1971},
  doi     = {10.1103/PhysRevA.4.2019}
}

@article{Dukelsky2004,
  author  = {Dukelsky, J. and Pittel, S. and Sierra, G.},
  title   = {Colloquium: Exactly Solvable Richardson-Gaudin Models for Many-Body Quantum Systems},
  journal = {Reviews of Modern Physics},
  volume  = {76},
  pages   = {643--662},
  year    = {2004},
  doi     = {10.1103/RevModPhys.76.643}
}

@article{Haldane1983,
  author  = {Haldane, F. D. M.},
  title   = {Fractional Quantization of the Hall Effect: A Hierarchy of Incompressible Quantum Fluid States},
  journal = {Physical Review Letters},
  volume  = {51},
  pages   = {605--608},
  year    = {1983},
  doi     = {10.1103/PhysRevLett.51.605}
}

@article{TrugmanKivelson1985,
  author  = {Trugman, S. A. and Kivelson, S.},
  title   = {Exact Results for the Fractional Quantum Hall Effect with General Interactions},
  journal = {Physical Review B},
  volume  = {31},
  pages   = {5280--5284},
  year    = {1985},
  doi     = {10.1103/PhysRevB.31.5280}
}

@article{RokhsarKivelson1988,
  author  = {Rokhsar, Daniel S. and Kivelson, Steven A.},
  title   = {Superconductivity and the Quantum Hard-Core Dimer Gas},
  journal = {Physical Review Letters},
  volume  = {61},
  pages   = {2376--2379},
  year    = {1988},
  doi     = {10.1103/PhysRevLett.61.2376}
}

@article{GirvinMacDonaldPlatzman1985,
  author  = {Girvin, S. M. and MacDonald, A. H. and Platzman, P. M.},
  title   = {Collective-Excitation Gap in the Fractional Quantum Hall Effect},
  journal = {Physical Review Letters},
  volume  = {54},
  pages   = {581--583},
  year    = {1985},
  doi     = {10.1103/PhysRevLett.54.581}
}

@article{NancarrowXin2023,
  author  = {Nancarrow, Colin Oscar and Xin, Yuan},
  title   = {Bootstrapping the Gap in Quantum Spin Systems},
  journal = {Journal of High Energy Physics},
  volume  = {2023},
  number  = {8},
  pages   = {052},
  year    = {2023},
  doi     = {10.1007/JHEP08(2023)052},
  eprint  = {2211.03819},
  archivePrefix = {arXiv}
}

@article{GaoHanKhalaf2026,
  author  = {Gao, Qiang and Han, Zhaoyu and Khalaf, Eslam},
  title   = {Bootstrapping Flatband Superconductors: Rigorous Lower Bounds on Superfluid Stiffness},
  journal = {Physical Review Letters},
  volume  = {136},
  pages   = {076503},
  year    = {2026},
  doi     = {10.1103/gw85-5r92},
  eprint  = {2506.18969},
  archivePrefix = {arXiv}
}

@article{Caux2009,
  author  = {Caux, Jean-S{\'e}bastien},
  title   = {Correlation Functions of Integrable Models: A Description of the {ABACUS} Algorithm},
  journal = {Journal of Mathematical Physics},
  volume  = {50},
  pages   = {095214},
  year    = {2009},
  doi     = {10.1063/1.3216474},
  eprint  = {0908.1660},
  archivePrefix = {arXiv}
}

@article{MurgKorepinVerstraete2012,
  author  = {Murg, V. and Korepin, V. E. and Verstraete, F.},
  title   = {Algebraic Bethe Ansatz and Tensor Networks},
  journal = {Physical Review B},
  volume  = {86},
  pages   = {045125},
  year    = {2012},
  doi     = {10.1103/PhysRevB.86.045125}
}

@article{AKLT1987,
  author  = {Affleck, Ian and Kennedy, Tom and Lieb, Elliott H. and Tasaki, Hal},
  title   = {Rigorous Results on Valence-Bond Ground States in Antiferromagnets},
  journal = {Physical Review Letters},
  volume  = {59},
  pages   = {799--802},
  year    = {1987},
  doi     = {10.1103/PhysRevLett.59.799}
}

@article{DeTomasiHetterichSalaPollmann2019,
  author  = {De Tomasi, Giuseppe and Hetterich, Daniel and Sala, Pablo and Pollmann, Frank},
  title   = {Dynamics of Strongly Interacting Systems: From Fock-Space Fragmentation to Many-Body Localization},
  journal = {Physical Review B},
  volume  = {100},
  pages   = {214313},
  year    = {2019},
  doi     = {10.1103/PhysRevB.100.214313}
}

@article{Tovmasyan2016,
  author  = {Tovmasyan, Murad and Peotta, Sebastiano and T{\"o}rm{\"a}, P{\"a}ivi and Huber, Sebastian D.},
  title   = {Effective Theory and Emergent {SU}(2) Symmetry in the Flat Bands of Attractive Hubbard Models},
  journal = {Physical Review B},
  volume  = {94},
  pages   = {245149},
  year    = {2016},
  doi     = {10.1103/PhysRevB.94.245149}
}

@article{Liang2017,
  author  = {Liang, Long and Vanhala, Tuomas I. and Peotta, Sebastiano and Siro, Topi and Harju, Ari and T{\"o}rm{\"a}, P{\"a}ivi},
  title   = {Band Geometry, Berry Curvature, and Superfluid Weight},
  journal = {Physical Review B},
  volume  = {95},
  pages   = {024515},
  year    = {2017},
  doi     = {10.1103/PhysRevB.95.024515}
}

@article{TormaLiangPeotta2018,
  author  = {T{\"o}rm{\"a}, P{\"a}ivi and Liang, Long and Peotta, Sebastiano},
  title   = {Quantum Metric and Effective Mass of a Two-Body Bound State in a Flat Band},
  journal = {Physical Review B},
  volume  = {98},
  pages   = {220511},
  year    = {2018},
  doi     = {10.1103/PhysRevB.98.220511}
}

@article{Sondhi1993,
  author  = {Sondhi, S. L. and Karlhede, A. and Kivelson, S. A. and Rezayi, E. H.},
  title   = {Skyrmions and the Crossover from the Integer to Fractional Quantum Hall Effect at Small Zeeman Energies},
  journal = {Physical Review B},
  volume  = {47},
  pages   = {16419--16426},
  year    = {1993},
  doi     = {10.1103/PhysRevB.47.16419}
}

@article{Moon1995,
  author  = {Moon, K. and Mori, H. and Yang, Kun and Girvin, S. M. and MacDonald, A. H. and Zheng, L. and Yoshioka, D. and Zhang, Shou-Cheng},
  title   = {Spontaneous Interlayer Coherence in Double-Layer Quantum Hall Systems: Charged Vortices and Kosterlitz-Thouless Phase Transitions},
  journal = {Physical Review B},
  volume  = {51},
  pages   = {5138--5170},
  year    = {1995},
  doi     = {10.1103/PhysRevB.51.5138}
}

@article{WuDasSarmaQAHF2020,
  author  = {Wu, Fengcheng and Das Sarma, Sankar},
  title   = {Collective Excitations of Quantum Anomalous Hall Ferromagnets in Twisted Bilayer Graphene},
  journal = {Physical Review Letters},
  volume  = {124},
  pages   = {046403},
  year    = {2020},
  doi     = {10.1103/PhysRevLett.124.046403}
}

@article{WuDasSarma2020,
  author  = {Wu, Fengcheng and Das Sarma, S.},
  title   = {Quantum Geometry and Stability of Moir{\'e} Flatband Ferromagnetism},
  journal = {Physical Review B},
  volume  = {102},
  pages   = {165118},
  year    = {2020},
  doi     = {10.1103/PhysRevB.102.165118}
}

@article{RepellinDongZhangSenthil2020,
  author  = {Repellin, C{\'e}cile and Dong, Zhihuan and Zhang, Ya-Hui and Senthil, T.},
  title   = {Ferromagnetism in Narrow Bands of Moir{\'e} Superlattices},
  journal = {Physical Review Letters},
  volume  = {124},
  pages   = {187601},
  year    = {2020},
  doi     = {10.1103/PhysRevLett.124.187601}
}

@article{HuHyartPikulinRossi2022,
  author  = {Hu, Xiang and Hyart, Timo and Pikulin, Dmitry I. and Rossi, Enrico},
  title   = {Quantum-Metric-Enabled Exciton Condensate in Double Twisted Bilayer Graphene},
  journal = {Physical Review B},
  volume  = {105},
  pages   = {L140506},
  year    = {2022},
  doi     = {10.1103/PhysRevB.105.L140506}
}

@article{VermaGuerciQueiroz2024,
  author  = {Verma, Nishchhal and Guerci, Daniele and Queiroz, Raquel},
  title   = {Geometric Stiffness in Interlayer Exciton Condensates},
  journal = {Physical Review Letters},
  volume  = {132},
  pages   = {236001},
  year    = {2024},
  doi     = {10.1103/PhysRevLett.132.236001}
}

@article{MaoChowdhury24,
  author  = {Mao, Dan and Chowdhury, Debanjan},
  title   = {Upper Bounds on Superconducting and Excitonic Phase Stiffness for Interacting Isolated Narrow Bands},
  journal = {Physical Review B},
  volume  = {109},
  pages   = {024507},
  year    = {2024},
  doi     = {10.1103/PhysRevB.109.024507}
}

@article{TBGV,
  author  = {Bernevig, B. Andrei and Lian, Biao and Cowsik, Aditya and Xie, Fang and Regnault, Nicolas and Song, Zhi-Da},
  title   = {Twisted Bilayer Graphene. V. Exact Analytic Many-Body Excitations in Coulomb Hamiltonians: Charge Gap, Goldstone Modes, and Absence of Cooper Pairing},
  journal = {Physical Review B},
  volume  = {103},
  pages   = {205415},
  year    = {2021},
  doi     = {10.1103/PhysRevB.103.205415},
  eprint  = {2009.14200},
  archivePrefix = {arXiv}
}

@article{SchindlerTrion2022,
  author  = {Schindler, Frank and Vafek, Oskar and Bernevig, B. Andrei},
  title   = {Trions in Twisted Bilayer Graphene},
  journal = {Physical Review B},
  volume  = {105},
  pages   = {155135},
  year    = {2022},
  doi     = {10.1103/PhysRevB.105.155135},
  eprint  = {2112.12776},
  archivePrefix = {arXiv}
}

@article{KhalafVishwanath2022,
  author  = {Khalaf, Eslam and Vishwanath, Ashvin},
  title   = {Baby Skyrmions in Chern Ferromagnets and Topological Mechanism for Spin-Polaron Formation in Twisted Bilayer Graphene},
  journal = {Nature Communications},
  volume  = {13},
  pages   = {6245},
  year    = {2022},
  doi     = {10.1038/s41467-022-33673-3}
}

@article{ZhangSunLiPanMeng2022,
  author  = {Zhang, Xu and Sun, Kai and Li, Heqiu and Pan, Gaopei and Meng, Zi Yang},
  title   = {Superconductivity and Bosonic Fluid Emerging from Moir{\'e} Flat Bands},
  journal = {Physical Review B},
  volume  = {106},
  pages   = {184517},
  year    = {2022},
  doi     = {10.1103/PhysRevB.106.184517}
}

@article{ChanGremeaudBatrouni2022Designer,
  author  = {Chan, Si Min and Gr{\'e}maud, B. and Batrouni, G. G.},
  title   = {Designer Flat Bands: Topology and Enhancement of Superconductivity},
  journal = {Physical Review B},
  volume  = {106},
  pages   = {104514},
  year    = {2022},
  doi     = {10.1103/PhysRevB.106.104514}
}

@article{KhemaniHermeleNandkishore2020,
  author  = {Khemani, Vedika and Hermele, Michael and Nandkishore, Rahul},
  title   = {Localization from Hilbert Space Shattering: From Theory to Physical Realizations},
  journal = {Physical Review B},
  volume  = {101},
  pages   = {174204},
  year    = {2020},
  doi     = {10.1103/PhysRevB.101.174204},
  eprint  = {1910.01137},
  archivePrefix = {arXiv}
}

@article{MoudgalyaMotrunich2022,
  author  = {Moudgalya, Sanjay and Motrunich, Olexei I.},
  title   = {Hilbert Space Fragmentation and Commutant Algebras},
  journal = {Physical Review X},
  volume  = {12},
  pages   = {011050},
  year    = {2022},
  doi     = {10.1103/PhysRevX.12.011050},
  eprint  = {2108.10324},
  archivePrefix = {arXiv}
}

@misc{HanHartKhudorozhkovNandkishore2026,
  author        = {Han, Yiqiu and Hart, Oliver and Khudorozhkov, Alexey and Nandkishore, Rahul},
  title         = {Quantum Fragmentation},
  year          = {2026},
  eprint        = {2604.06461},
  archivePrefix = {arXiv},
  primaryClass  = {cond-mat.stat-mech}
}

@article{PT15,
  author  = {Peotta, Sebastiano and T{\"o}rm{\"a}, P{\"a}ivi},
  title   = {Superfluidity in topologically nontrivial flat bands},
  journal = {Nature Communications},
  volume  = {6},
  pages   = {8944},
  year    = {2015},
  doi     = {10.1038/ncomms9944}
}

@article{Julku16,
  author  = {Julku, Aleksi and Peotta, Sebastiano and Vanhala, Tuomas I. and Kim, Dong-Hee and T{\"o}rm{\"a}, P{\"a}ivi},
  title   = {Geometric origin of superfluidity in the {Lieb}-lattice flat band},
  journal = {Phys. Rev. Lett.},
  volume  = {117},
  pages   = {045303},
  year    = {2016},
  doi     = {10.1103/PhysRevLett.117.045303}
}

@article{TPB22,
  author  = {T{\"o}rm{\"a}, P{\"a}ivi and Peotta, Sebastiano and Bernevig, Bogdan A.},
  title   = {Superconductivity, superfluidity and quantum geometry in twisted multilayer systems},
  journal = {Nature Reviews Physics},
  volume  = {4},
  pages   = {528--542},
  year    = {2022},
  doi     = {10.1038/s42254-022-00466-y}
}

@article{HA22,
  author        = {Herzog-Arbeitman, Jonah and Chew, Aaron and Huhtinen, Kukka-Emilia and T{\"o}rm{\"a}, P{\"a}ivi and Bernevig, B. Andrei},
  title         = {Many-body superconductivity in topological flat bands},
  journal       = {Communications Physics},
  year          = {2026},
  doi           = {10.1038/s42005-026-02732-2},
  eprint        = {2209.00007},
  archivePrefix = {arXiv}
}

@article{Huhtinen22,
  author  = {Huhtinen, Kukka-Emilia and Herzog-Arbeitman, Jonah and Chew, Aaron and Bernevig, B. Andrei and T{\"o}rm{\"a}, P{\"a}ivi},
  title   = {Revisiting flat band superconductivity: Dependence on minimal quantum metric and band touchings},
  journal = {Phys. Rev. B},
  volume  = {106},
  pages   = {014518},
  year    = {2022},
  doi     = {10.1103/PhysRevB.106.014518}
}

@article{QGN,
  author  = {Han, Zhaoyu and Herzog-Arbeitman, Jonah and Bernevig, B. Andrei and Kivelson, Steven A.},
  title   = {``Quantum Geometric Nesting'' and Solvable Model Flat-Band Systems},
  journal = {Phys. Rev. X},
  volume  = {14},
  pages   = {041004},
  year    = {2024},
  doi     = {10.1103/PhysRevX.14.041004}
}

@article{HanKivelson25,
  author  = {Han, Zhaoyu and Kivelson, Steven A.},
  title   = {Models of interacting bosons with exact ground states: A unified approach},
  journal = {Phys. Rev. B},
  volume  = {111},
  pages   = {174520},
  year    = {2025},
  doi     = {10.1103/PhysRevB.111.174520},
  eprint  = {2408.15319},
  archivePrefix = {arXiv}
}

@article{Chiral,
  author        = {Han, Zhaoyu and Herzog-Arbeitman, Jonah and Gao, Qiang and Khalaf, Eslam},
  title         = {Exact models of chiral flat-band superconductors},
  journal       = {arXiv preprint},
  eprint        = {2508.21127},
  archivePrefix = {arXiv},
  year          = {2025},
  note          = {arXiv:2508.21127}
}

@article{SWZ93,
  author  = {Scalapino, Douglas J. and White, Steven R. and Zhang, Shoucheng},
  title   = {Insulator, metal, or superconductor: The criteria},
  journal = {Phys. Rev. B},
  volume  = {47},
  pages   = {7995--8007},
  year    = {1993},
  doi     = {10.1103/PhysRevB.47.7995}
}

@article{PTR98,
  author  = {Paramekanti, Arun and Trivedi, Nandini and Randeria, Mohit},
  title   = {Upper bounds on the superfluid stiffness of disordered systems},
  journal = {Phys. Rev. B},
  volume  = {57},
  pages   = {11639--11647},
  year    = {1998},
  doi     = {10.1103/PhysRevB.57.11639}
}

@article{HVR19,
  author  = {Hazra, Tamaghna and Verma, Nishchhal and Randeria, Mohit},
  title   = {Bounds on the Superconducting Transition Temperature: Applications to Twisted Bilayer Graphene and Cold Atoms},
  journal = {Phys. Rev. X},
  volume  = {9},
  pages   = {031049},
  year    = {2019},
  doi     = {10.1103/PhysRevX.9.031049}
}

@article{VHR21,
  author  = {Verma, Nishchhal and Hazra, Tamaghna and Randeria, Mohit},
  title   = {Optical spectral weight, phase stiffness, and $T_c$ bounds for trivial and topological flat band superconductors},
  journal = {Proceedings of the National Academy of Sciences},
  volume  = {118},
  pages   = {e2106744118},
  year    = {2021},
  doi     = {10.1073/pnas.2106744118}
}

@article{Xie20,
  author  = {Xie, Fang and Song, Zhida and Lian, Biao and Bernevig, B. Andrei},
  title   = {Topology-Bounded Superfluid Weight in Twisted Bilayer Graphene},
  journal = {Phys. Rev. Lett.},
  volume  = {124},
  pages   = {167002},
  year    = {2020},
  doi     = {10.1103/PhysRevLett.124.167002}
}

@article{HAPeri22,
  author  = {Herzog-Arbeitman, Jonah and Peri, Valerio and Schindler, Frank and Huber, Sebastian D. and Bernevig, B. Andrei},
  title   = {Superfluid Weight Bounds from Symmetry and Quantum Geometry in Flat Bands},
  journal = {Phys. Rev. Lett.},
  volume  = {128},
  pages   = {087002},
  year    = {2022},
  doi     = {10.1103/PhysRevLett.128.087002}
}

@article{MaoChowdhury23,
  author  = {Mao, Dan and Chowdhury, Debanjan},
  title   = {Diamagnetic response and phase stiffness for interacting isolated narrow bands},
  journal = {Proceedings of the National Academy of Sciences},
  volume  = {120},
  pages   = {e2217816120},
  year    = {2023},
  doi     = {10.1073/pnas.2217816120}
}

@article{PhysRevLett.133.176001,
  title = {Critical Spontaneous Breaking of U(1) Symmetry at Zero Temperature in One Dimension},
  author = {Watanabe, Haruki and Katsura, Hosho and Lee, Jong Yeon},
  journal = {Phys. Rev. Lett.},
  volume = {133},
  issue = {17},
  pages = {176001},
  numpages = {6},
  year = {2024},
  month = {Oct},
  publisher = {American Physical Society},
  doi = {10.1103/PhysRevLett.133.176001},
  url = {https://link.aps.org/doi/10.1103/PhysRevLett.133.176001}
}

@article{AlaviradSau2020,
  author  = {Alavirad, Yahya and Sau, Jay D.},
  title   = {Ferromagnetism and Its Stability from the One-Magnon Spectrum in Twisted Bilayer Graphene},
  journal = {Physical Review B},
  volume  = {102},
  pages   = {235123},
  year    = {2020},
  doi     = {10.1103/PhysRevB.102.235123},
  eprint  = {1907.13633},
  archivePrefix = {arXiv}
}

@article{PhysRevB.110.165109,
  title = {Quantum fragmentation in the extended quantum breakdown model},
  author = {Chen, Bo-Ting and Prem, Abhinav and Regnault, Nicolas and Lian, Biao},
  journal = {Phys. Rev. B},
  volume = {110},
  issue = {16},
  pages = {165109},
  numpages = {26},
  year = {2024},
  month = {Oct},
  publisher = {American Physical Society},
  doi = {10.1103/PhysRevB.110.165109},
  url = {https://link.aps.org/doi/10.1103/PhysRevB.110.165109}
}

@misc{KhalafSoftModes2020,
  author        = {Khalaf, Eslam and Bultinck, Nick and Vishwanath, Ashvin and Zaletel, Michael P.},
  title         = {Soft Modes in Magic Angle Twisted Bilayer Graphene},
  year          = {2020},
  eprint        = {2009.14827},
  archivePrefix = {arXiv},
  primaryClass  = {cond-mat.str-el}
}

@article{KallinHalperin1984,
  author  = {Kallin, Catherine and Halperin, B. I.},
  title   = {Excitations from a Filled Landau Level in the Two-Dimensional Electron Gas},
  journal = {Physical Review B},
  volume  = {30},
  pages   = {5655--5668},
  year    = {1984},
  doi     = {10.1103/PhysRevB.30.5655}
}

@article{Rai2026hierarchyofspectral,
  doi = {10.22331/q-2026-04-13-2065},
  url = {https://doi.org/10.22331/q-2026-04-13-2065},
  title = {A {H}ierarchy of {S}pectral {G}ap {C}ertificates for {F}rustration-{F}ree {S}pin {S}ystems},
  author = {Rai, Kshiti Sneh and Kull, Ilya and Emonts, Patrick and Tura, Jordi and Schuch, Norbert and Baccari, Flavio},
  journal = {{Quantum}},
  issn = {2521-327X},
  publisher = {{Verein zur F{\"{o}}rderung des Open Access Publizierens in den Quantenwissenschaften}},
  volume = {10},
  pages = {2065},
  month = apr,
  year = {2026}
}

@article{PhysRevB.110.195140,
  title = {Quadratic dispersion relations in gapless frustration-free systems},
  author = {Masaoka, Rintaro and Soejima, Tomohiro and Watanabe, Haruki},
  journal = {Phys. Rev. B},
  volume = {110},
  issue = {19},
  pages = {195140},
  numpages = {8},
  year = {2024},
  month = {Nov},
  publisher = {American Physical Society},
  doi = {10.1103/PhysRevB.110.195140},
  url = {https://link.aps.org/doi/10.1103/PhysRevB.110.195140}
}

@article{WuLiYao,
    author = {Wu, Zhengzhi and Li, Ming-Rui and Yao, Hong},
  journal = {To appear}
}

@article{annurev:/content/journals/10.1146/annurev-conmatphys-031119-050644,
   author = "Watanabe, Haruki",
   title = "Counting Rules of Nambu–Goldstone Modes", 
   journal= "Annual Review of Condensed Matter Physics",
   year = "2020",
   volume = "11",
   number = "Volume 11, 2020",
   pages = "169-187",
   doi = "https://doi.org/10.1146/annurev-conmatphys-031119-050644",
   url = "https://www.annualreviews.org/content/journals/10.1146/annurev-conmatphys-031119-050644",
   publisher = "Annual Reviews",
   issn = "1947-5462",
   type = "Journal Article",
  }

@article{PhysRevX.4.031057,
  title = {Effective Lagrangian for Nonrelativistic Systems},
  author = {Watanabe, Haruki and Murayama, Hitoshi},
  journal = {Phys. Rev. X},
  volume = {4},
  issue = {3},
  pages = {031057},
  numpages = {36},
  year = {2014},
  month = {Sep},
  publisher = {American Physical Society},
  doi = {10.1103/PhysRevX.4.031057},
  url = {https://link.aps.org/doi/10.1103/PhysRevX.4.031057}
}

@article{PhysRevB.105.024502,
  title = {Pairing and superconductivity in quasi-one-dimensional flat-band systems: Creutz and sawtooth lattices},
  author = {Chan, Si Min and Gr\'emaud, B. and Batrouni, G. G.},
  journal = {Phys. Rev. B},
  volume = {105},
  issue = {2},
  pages = {024502},
  numpages = {10},
  year = {2022},
  month = {Jan},
  publisher = {American Physical Society},
  doi = {10.1103/PhysRevB.105.024502},
  url = {https://link.aps.org/doi/10.1103/PhysRevB.105.024502}
}

@article{PhysRevLett.130.226001,
  title = {Superconductivity, Charge Density Wave, and Supersolidity in Flat Bands with a Tunable Quantum Metric},
  author = {Hofmann, Johannes S. and Berg, Erez and Chowdhury, Debanjan},
  journal = {Phys. Rev. Lett.},
  volume = {130},
  issue = {22},
  pages = {226001},
  numpages = {6},
  year = {2023},
  month = {May},
  publisher = {American Physical Society},
  doi = {10.1103/PhysRevLett.130.226001},
  url = {https://link.aps.org/doi/10.1103/PhysRevLett.130.226001}
}

@article{PhysRevB.102.201112,
  title = {Superconductivity, pseudogap, and phase separation in topological flat bands},
  author = {Hofmann, Johannes S. and Berg, Erez and Chowdhury, Debanjan},
  journal = {Phys. Rev. B},
  volume = {102},
  issue = {20},
  pages = {201112(R)},
  numpages = {6},
  year = {2020},
  month = {Nov},
  publisher = {American Physical Society},
  doi = {10.1103/PhysRevB.102.201112},
  url = {https://link.aps.org/doi/10.1103/PhysRevB.102.201112}
}

\end{document}